\documentclass[twocolumn]{aastex631}

\accepted{2024 November 19}
\submitjournal{ApJ}

\shorttitle{Potential Interior Structures and Habitability of Four Super-Earth Exoplanets}
\shortauthors{}
\graphicspath{{./}{figures/}}

\begin{document}

\title{Potential Interior Structures and Habitability of Super-Earth Exoplanets LHS\,1140\,b, K2-18\,b, TOI-1452\,b and TOI-1468\,c}

\correspondingauthor{Mangesh Daspute}
\email{mangeshd@ariel.ac.il}
\author[0009-0006-2517-3065]{Mangesh Daspute}
\affiliation{Department of Physics, Ariel University, Ariel 40700, Israel}


\author[0000-0002-8900-212X]{Amri Wandel}
\affiliation{The Racah Institute of Physics, The Hebrew University of Jerusalem, 91904, Israel}



\author[0000-0002-5893-2471]{Ravi Kumar Kopparapu}
\affiliation{NASA Goddard Space Flight Center, 8800 Greenbelt Road,
Greenbelt, MD 20771, USA}
\author[0000-0002-6859-0882]{Volker Perdelwitz}
\affiliation{Department of Earth and Planetary Sciences, Weizmann Institute of Science, Rehovot 76100, Israel;}

\author{Jerusalem Tamirat Teklu}
\affiliation{Department of Physics, Ariel University, Ariel 40700, Israel}
\author[0000-0003-3757-1440]{Lev Tal-Or}
\affiliation{Department of Physics, Ariel University, Ariel 40700, Israel}
\affiliation{Astrophysics, Geophysics And Space Science Research Center, Ariel University, Ariel 40700, Israel}



\begin{abstract}
We analyze four super-Earth exoplanets, LHS\,1140\,b, K2-18\,b, TOI-1452\,b, and TOI-1468\,c, which orbit M-dwarf stars in the habitable zone. Their relative proximity, within 40 parsecs, makes them prime candidates for follow-up observations and atmospheric and habitability studies. 
This paper aims to assess their internal structure and habitability, considering their tidal heating, atmospheric heating, and global transport. 
We model the interior structure of the planets by applying Bayesian inference to an exoplanet's interior model. 
A constant quality factor model is used to calculate the range of tidal heating, and a one-dimensional analytical model of tidally locked planets is used to assess their surface temperature distribution and habitability.
Assuming no or only thin atmospheres, K2-18\,b and TOI-1468\,c are likely to be water worlds. However, TOI-1452\,b and LHS\,1140\,b may have rocky surfaces. 
We find that tidal heating is not enough to raise the global mean surface temperature, but greenhouse heating can effectively do so. 
If the considered planets have retained thick atmospheres, K2-18\,b, TOI-1468\,c, and TOI-1452\,b may, for significant atmospheric heating and heat transport factors, be too hot to sustain liquid water on their surface. However, the lower instellation of LHS\,1140\,b and the non-zero probability of it having a rocky surface give more space for habitable conditions on the planet. 
\end{abstract}



\section{Introduction} 
\label{sec:intro}
As of 2024 September 19, there are 5756 confirmed exoplanets\footnote{NASA Exoplanet Archive: \url{exoplanetarchive.ipac.caltech.edu}}. 
Both mass and radius are known for 1297 of the exoplanets and 298 of these have a minimum mass between $1.9$\,M$_{\oplus}$ and $10.0$\,M$_{\oplus}$, also known as super-Earth exoplanets \citep{Charbonneau2009}. 
Unlike Earth, Super-Earth exoplanets potentially have thick ice layers \citep{Valencia_2007,Seager_2007} and could be scaled-up versions of icy satellites within our solar system like Europa or Ganymede \citep[e.g.,][]{1998Sci...281.2019A, 2004Sci...305..989A}. Depending upon the available water content during the formation of the super-Earths and their evolutionary history, they can have a significant amount of ice or water in them \citep{2004Icar..168....1R}. Transit and radial velocity techniques provide estimates of the bulk density of exoplanets based on their mass and radius measurements \citep{2000ApJ...532L..55M, Charbonneau2000, 2003ApJ...585.1038S} but knowing their interior structure just with remote observations is not possible. Different potential compositions that can explain the observed mass and radius can be found using a model of the interior structure of an exoplanet. Dozens of works looking into this topic were published in the last two decades \citep[e.g.,][]{Valencia_2007, Seager_2007, SOTIN_2007, Rogers_2010, Vazan2013, Unterborn_2016, Dorn2017, UnterbornEXOPLEX2018, vandenBerg2019, Boujibar2020, Acuna2021, Nixon2021, MagratheaHuang2022, HaldemannBiceps2024}. All of them emphasize the degenerate nature of the problem, since a range of different mass fractions and thicknesses of water-ice, mantle and core can explain the observed mass and radius. Nevertheless, the interior structure of an exoplanet is one of its most important aspects as it can influence its surface and atmospheric composition, in addition to habitability \citep{Gillmann2024}. The presence of water may also change the rheology of the mantle, reducing its effective viscosity by several orders of magnitude \citep{Katayama2008, Karato2015}.

Planets' habitability is a complex phenomenon that may be defined in several different ways. The classical habitable zone is defined as the circumstellar region in which a terrestrial mass planet ($0.3<$ M$_{p}<10\,M_{\oplus}$), under favourable atmospheric conditions, can sustain liquid water on its surface \citep{Huang1959A, Hart1978, Kasting1993, Underwood2003, Selsis2007A, Kaltenegger_2011, Kopparapu_2013}. Many different factors can influence the habitability of a planet. For example, being a condensable greenhouse gas, water itself may significantly affect the surface conditions and can even cause a runaway greenhouse effect \citep{Goldblatt2012}. For simplicity, we restrict ourselves here to the biohabitable zone, which is a domain in which liquid water can physically survive on at least part of the planetary surface \citep{Wandel_2018}. We do not consider complex geological or biological processes that may affect habitability, such as planetary magnetic fields that may protect its atmosphere and life on its surface \citep[e.g.,][]{Lundin2007, Chassefiere2007} or plate tectonic activity which drives the carbon-rock cycle and stabilizes temperature on the surface of a planet \citep[e.g.,][]{Alibert_2014}.


\begin{table*}[!t]
 \centering
 \setlength{\tabcolsep}{8pt} 
 \caption{Host-star and planet parameters.}
 \label{tab:table_properties}
 \begin{tabular}{l c c c c}
 \hline
 \hline
 Host star & LHS\,1140 & K2-18 & TOI-1452 & TOI-1468 \\ 
 \hline
 RA\,[hh:mm:ss] & 00:44:59.33\,$^{[2]}$ & 11:30:14.52\,$^{[2]}$ & 19:20:41.73\,$^{[2]}$ & 01:06:36.98\,$^{[2]}$ \\
 DEC\,[dd:mm:ss] & $-$15:16:17.54\,$^{[2]}$ & +07:35:18.26\,$^{[2]}$ & +73:11:43.54\,$^{[2]}$ & +19:13:33.16\,$^{[2]}$ \\ 
 $\pi$\,[mas] & $66.829 \pm 0.048^{[2]}$ & $26.247\pm 0.027^{[2]}$ & $32.782 \pm 0.014^{[2]}$ & $40.452 \pm 0.036^{[2]}$ \\ 
 $\mu_{\alpha}$\,[mas\,yr$^{-1}$] & $318.152\pm0.049^{[2]}$ & $-80.376\pm0.083^{[2]}$ & $7.800\pm0.017^{[2]}$ & $-42.067^{+0.047}_{-0.032}$\,$^{[2]}$ \\
 $\mu_{\delta}$\,[mas\,yr$^{-1}$] & $-596.623\pm0.054$\,$^{[2]}$ & $-133.142\pm0.063$\,$^{[2]}$ & $-74.076\pm0.017$\,$^{[2]}$ & $-222.790^{+0.047}_{-0.032}$\,$^{[2]}$ \\
 Distance\,[pc] & $14.9636\pm0.0107$\,$^{[2]}$ & $38.100\pm0.0385$\,$^{[2]}$ & $30.504\pm0.0130$\,$^{[2]}$ & $24.721\pm0.0219$\,$^{[2]}$\\
 G magnitude & $12.6539\pm0.0028$\,$^{[2]}$ & $12.4007\pm0.0028$\,$^{[2]}$ & $13.5982\pm0.0028$\,$^{[2]}$ & $12.104721\pm0.0028$\,$^{[2]}$ \\
 Spectral Type & M4.5V$^{[9]}$ & M3V$^{[10]}$ & M4V$^{[12]}$ & M3.0V$^{[13]}$ \\ 
 
 V magnitude & $14.150\pm0.06$\,$^{[8]}$ & $13.50\pm0.05$\,$^{[8]}$ & $14.354\pm0.121$\,$^{[2]}$ & $12.5\pm0.2$ \\
 L\,[L$_{sun}$] & $0.00398\pm0.00003$\,$^{[12]}$ & $0.0253\pm0.0021$\,$^{[3]}$ & $0.0070\pm 0.0006$\,$^{[12]}$ & $0.01595\pm0.0009$\,$^{[13]}$\\

 
Age\,[Gyr] & $>5$\,$^{[7]}$ & $2.4\pm0.6$\,$^{[18]}$ & & $1-10$\,$^{[16]}$ \\ 
 $T_\mathrm{eff}$\,[K] & $3096\pm48$\,$^{[12]}$ & $3457\pm39$\,$^{[3]}$ & $3185\pm50$\,$^{[12]}$ & $3496\pm25$\,$^{[13]}$ \\
 $R_{*}$\,[R$_{\sun}$] & $0.2159\pm0.0030$\,$^{[12]}$ & $0.4445\pm0.0148$\,$^{[3]}$ & $0.275\pm0.009$\,$^{[12]}$ & $0.344\pm0.005$\,$^{[13]}$ \\ 
 $M_{*}$\,[M$_{\sun}$] & $0.1844\pm0.0045$\,$^{[12]}$ & $0.4951\pm0.0043$\,$^{[3]}$ & $0.249\pm0.008$\,$^{[12]}$ & $0.339\pm0.011$\,$^{[13]}$ \\
$v\sin{i}$\,[km s$^{-1}$] &  & $<2$\,$^{[19]}$ & & $<2$\,$^{[6]}$\\
 $[\rm Fe/H]$\,$[\rm dex]$ & $-0.15\pm0.09$\,$^{[12]}$ & $0.12\pm0.16$\,$^{[11]}$ & $-0.07\pm0.02$\,$^{[12]}$ & $-0.040\pm0.070$\,$^{[14]}$ \\
 $\log \overline{R^\prime_{HK}}$ & $-5.740\pm0.006$\,$^{[15]}$ & $-5.055\pm0.031$\,$^{[15]}$ & & $-4.914\pm0.011$\,$^{[15]}$ \\
$P_\mathrm{rot}$\,[days] & $131\pm5$\,$^{[7]}$ & $39.63\pm0.50$\,$^{[11]}$ & $>120$\,$^{[12]}$ & $41$--$44$\,$^{[13]}$\\
 \hline
 Planet & LHS\,1140 b & K2-18\,b & TOI-1452\,b & TOI-1468\,c  \\ 
 \hline
 $M_{p}$\,[M$_{\oplus}$] & $5.60\pm0.19$\,$^{[12]}$ & $8.63\pm1.35$\,$^{[3]}$ & $4.82\pm1.30$\,$^{[5]}$ & $6.64^{+0.67}_{-0.68}$\,$^{[6]}$\\
 $R_{p}$\,[R$_{\oplus}$] & $1.730\pm0.025$\,$^{[12]}$ & $2.610\pm0.087$\,$^{[3]}$ & $1.672\pm0.071$\,$^{[5]}$ & $2.064\pm0.044$\,$^{[6]}$ \\ 
 Eccentricity & $<0.043$\,$^{[12]}$ & $0.20\pm0.08$\,$^{[17]}$ & $0$\,$^{[5]*}$ & $0$\,$^{[6]*}$ \\
 Semimajor axis\,[AU] & $0.0946\pm0.0017$\,$^{[12]}$ & $0.1429^{+0.0060}_{-0.0065}$\,$^{[3]}$ & $0.061\pm0.003$\,$^{[5]}$ & $0.0859^{+0.0013}_{-0016}$\,$^{[6]}$ \\
 Instellation\,[S$_{\oplus}$] & $0.43\pm0.03$\,$^{[12]}$ & $1.005^{+0.084}_{-0.079}$\,$^{[3]}$ & $1.8\pm0.2$\,$^{[5]}$ & $2.15\pm0.09$\,$^{[6]}$ \\ 
 $T_\mathrm{eq}$\,[K] (Albedo=0)& $226\pm4$\,$^{[12]}$ & $254\pm3.9$\,$^{[3]}$ & $326\pm7$\,$^{[5]}$ & $337.5^{+3.7}_{-3.4}$\,$^{[6]}$\\
 $\rho$\,[g\,cm$^{-3}$] & $5.9\pm0.3$\,$^{[12]}$ & $2.67^{+0.52}_{-0.47}$\,$^{[3]}$ & $5.6^{+1.8}_{-1.6}$\,$^{[5]}$ & $4.15^{+0.69}_{-0.68}$\,$^{[6]}$\\
 g\,[m\,s$^{-2}$] & $18.33\pm0.63$\,$^{[12]}$ & $12.43^{+2.17}_{-2.07}$\,$^{[3]}$ & $16.90\pm5.99$\,$^{[5]}$ & $15.26^{+1.68}_{-1.63}$\,$^{[6]}$\\
$v_\mathrm{escape}$\,[km\,s$^{-1}$] & $34.83\pm0.84^{[15]}$ & $28.77\pm2.73^{[15]}$ & $26.86\pm4.19^{[15]}$ & $28.40\pm1.74^{[15]}$ \\
Orbital Period\,[days] & $24.73723\pm0.00002$\,$^{[12]}$ & $32.939623^{+0.000095}_{-0.000100}$\,$^{[3]}$ & $11.06201\pm0.00002$\,$^{[5]}$ & $15.532482^{+0.000034}_{-0.000033}$\,$^{[6]}$ \\
\hline
\end{tabular}\\
 
Notes. $^{*}$ Keplerian model with eccentric orbit was only marginally favoured over circular orbit, so eccentricity was fixed to 0 during joint RV and transit analysis in \citet{chaturvedi_2022} and \citet{Cadieux_2022}\\

References. (1) \citet{Gaia_Collaboration_2018}; (2) \citet{Gaia_Collaboration_2020}; (3) \citet{Benneke_2019}; (4) \citet{Lillo-Box_2020}; (5) \citet{Cadieux_2022}; (6) \citet{chaturvedi_2022}; (7) \citet{Dittmann_2017}; (8) \citet{Zacharis_2012_catalogue}; (9) \citet{Reid_1995}; (10) \citet{Dressing_2019}; (11) \citet{Sarkis_2018}; (12) \citet{Cadieux_2024}; (13) \citet{chaturvedi_2022}; (14) \citet{Marfil_2021};
(15) This work; (16) \citet{2022ApJS..258....9W}; (17) \citet{Sarkis_2018}; (18) \citet{Guinan_2019}; (19) \citet{Reiners2018}.

\end{table*}


Tidal forces arising from the host star and neighbouring planets can alter the orbital properties and interior structure of a planet \citep{Henning_2009, Fabrycky2009}. In particular, potentially habitable exoplanets around M-dwarf stars may be in synchronous rotation, which can result in one side of the planet always facing the star with perpetual daylight \citep[e.g.,][]{2017CeMDA.129..509B}. This allows for biohabitable conditions beyond the classical habitable zone \citep{Yang_2013, Wandel_2018}. To find the best candidates for future targeted observations, for instance with the \textit{JWST} \citep{GardnerJWST2006}, it is important to assess the tidally locked classical habitable exoplanets considering their interior, tidal heating, and atmospheric properties. Here, we focus on super-Earth exoplanets orbiting M-dwarf stars, which have radius measurement and instellation between 0.1 and 2.5 S$_{\oplus}$. We found four exoplanets that meet these criteria, namely LHS\,1140\,b, K2-18\,b, TOI-1452\,b, and TOI-1468\,c. All four are super-Earth exoplanets orbiting M-dwarf stars in their classical habitable zones \citep{Dittmann_2017, Montet_2015, Cadieux_2022, chaturvedi_2022}, which also locates them inside the tidal locking region of their host stars. All four stars are within a distance of 40 parsec \citep{Gaia_Collaboration_2020}. Having four similar planets with different bulk densities and equilibrium temperatures, enabled us to perform a comparative study.

Recent studies could place constrains on the atmospheric conditions in two of the four planets. \textit{HST} observations of LHS\,1140\,b show marginal evidence of water in its atmosphere, but the primary or secondary nature of the atmosphere cannot be determined \citep{Edwards_2021}. \citet{Cadieux_JWST_2024} analyzed the 0.65–2.7 $\mu$m transmission spectrum of LHS\,1140\,b obtained from two visits with \textit{JWST}/NIRISS. \citet{Damiano_2024} analyzed two \textit{JWST}/NIRSpec spectra between 1.7 and 5.2 $\mu$m. Both studies exclude a clear H2-rich atmosphere for LHS\,1140\,b, with the most likely atmospheric scenario being that of a thin N$_2$ or CO$_2$ dominated atmosphere. Recently, a \textit{JWST} study of K2-18\,b detected CH$_4$ and CO$_2$ in a H$_2$ rich atmosphere \citep{Madhusudhan_2023}. It supports a hycean nature of the exoplanet \citep{Madhusudhan_2021}, meaning that the planet allows for large oceans with habitable conditions underneath a H$_2$ rich atmosphere. However, there was no significant evidence for H$_2$O, NH$_3$, CO, or HCN in the atmosphere. \citet{Gomes2020} studied the spin and orbit evolution of exoplanetary systems K2-18 and LHS\,1140 considering tidal response and found that if the eccentricities of LHS\,1140\,b and LHS\,1140\,c are close to zero, then both are likely rocky, otherwise both are likely mini-Neptunes. Hence, a more precise estimation of their eccentricities may improve their characterisation using tidal response. TOI-1452 b and TOI-1468 c are newly discovered exoplanets, which were discovered by using \textit{TESS} photometry \citep{RickerTESS2014} and confirmed by using radial velocity measurements from ground-based spectrographs. 

Table \ref{tab:table_properties} gives the main properties of the four exoplanets and their host stars, as they were collected from the available literature. This includes equilibrium temperature considering zero albedo. $\log \overline{R^\prime_{HK}}$ for LHS\,1140 and K2-18 were derived by co-adding all available HARPS \citep{2003Msngr.114...20M} spectra and applying the extraction method introduced by \citet{2021A&A...652A.116P, Perdelwitz2024}. For TOI-1468, the single available HIRES \citep{1994SPIE.2198..362V} spectrum was analyzed in the same manner. No public spectrum in Ca H\&K wavelength was available for TOI-1452.

This paper has three main objectives: (a) to study the planets' interior structure and surface properties by applying Bayesian inference to their observed masses and radii, (b) to assess the contribution of tidal heating to the total heat budget of the planets by using their stellar and planetary properties, and (c) to assess their biohabitability by using the one-dimensional analytical model of \citet{Wandel_2018}. Instead of an in-detail study of either the interior structure or the atmospheric habitability of a single planet, we focus on a comparative study of both the interior structure and biohabitability of four similar planets. Consequently, simplistic one-dimensional models were preferred over sophisticated multi-parametric models. Nevertheless, this choice enabled us estimating the relative probability of the planetary surfaces being rocky or water rich, and assessing their potential biohabitability. In section \ref{sec:models} we describe our models of the interior structure, tidal heating, and biohabitability. 
The results for the four exoplanets studied here are presented in section \ref{sec:results}. 
We summarize and discuss our main findings in section \ref{sec:discussion}.

\section{Models and Methods} \label{sec:models}
In this section, we explain our modelling of the interior structure, equations of state, tidal heating, and biohabitability.

\subsection{Internal Structure}

In determining the internal structure of an exoplanet by finding the masses or sizes of the core, mantle, and ice layers, there are three unknowns and two constraints---the measured mass and radius of the exoplanet. Hence, it is an under-constrained problem. Additionally, uncertainties in the observed mass and radius make it difficult to precisely determine the interior structure of an exoplanet \citep{Dorn2015}. To mitigate this problem, we used constraints based on planet formation conditions and relative elemental abundance in the host star, as explained in \cite{Valencia_2007}. The relative elemental abundance of the host star can be a proxy for that of its planets \citep{lodders2003}. These constraints allow us to reject some compositions, i.e.\,the set of values of the planets' Core Mass Fraction (CMF), Mantle Mass Fraction (MMF), and Ice Mass Fraction (IMF), which are not supported by its host star relative elemental abundances. Furthermore, we used Bayesian inference to estimate the posterior probability density function (PDF) of the sizes and mass fractions of each layer inside each of the four exoplanets studied here.

In the last two decades there was a significant progress in planets' interior structure models and the equations of states, used by the models, got better at representing and assessing Earth-like and super-Earth compositions. Today, different interior structure models, which use different computational techniques and different equations of state, are in use \citep{Valencia_2007, Seager_2007, SOTIN_2007, Rogers_2010, Vazan2013, Unterborn_2016, Dorn2017, UnterbornEXOPLEX2018,vandenBerg2019, Boujibar2020, Acuna2021, Nixon2021, MagratheaHuang2022, HaldemannBiceps2024}. Here, we used MAGRATHEA \citep{MagratheaHuang2022}, which is an open-source interior solver to assess the interior structure of the four exoplanets. MAGRATHEA uses the latest developments in the equations of state \citep[e.g.,][]{Grande2022} from high-pressure physics experiments, allows users to modify planet models as needed and is under active development.
To enable a solution, we assume a completely differentiated radially symmetric planet in hydrostatic equilibrium.





\subsubsection{Equations of State} \label{sec:EoS}
Inside a planet, the density of a material at pressure P and temperature T is determined by its Equation Of State (EOS). For instance, the pressure at the Earth's core is about $330$\,GPa. Super-Earths have higher pressures in their cores. To describe the relationship between pressure, density and temperature of the core, mantle, ice, and atmosphere, we used the default EOS described in section 3.3 of \citet{MagratheaHuang2022}. The default EOS feature up-to-date experimental results from high-pressure physics. Although FeS and FeO alloys in the core are not included, the core is represented by the Vinet fit from \citet[][Eq. 9]{Smith2018}. The upper mantle is represented by silicate perovskite at lower temperatures using the EOS given in \citet{Oganov2004}. Post-perovskite EOS in the lower mantle is based on \citet{sakai2016experimental}. Hydrosphere is represented by Ice VII with thermal expansion based on \citet{Bezacier2014}, which transitions to a Vinet EOS for Ice X from \citet{Grande2022} at 30.9 GPa. MAGRATHEA considers an isothermal atmosphere at a pressure below 100 bar and an adiabatic temperature gradient above 100 bar.



 
\subsubsection{Constraints from Relative Elemental Abundance}

Most of the stars in the solar neighbourhood have close to solar relative elemental abundance, which is considered a good proxy for the bulk composition of their planets \citep{Gilli2006, Grasset2009, Dorn2015, Adibekyan2015, Zeng2016, Unterborn_2016, Hinkel_2018}. Exoplanet interiors cannot consist of random mass fractions of core, mantle, and water-ice layers. We have used constraints based on planet formation conditions and the relative elemental abundance of the host star explained in \citet{Valencia_2007}. These constraints allow us to reject some compositions, i.e. the set of values of CMF, MMF, and IMF, which are improbable considering planets form in the protoplanetary disc of the host star. The model assumes the metallicity of the host star to be similar to the Sun, which is approximately true for the exoplanets we analyze here. Table \ref{tab:table_properties} shows that the metallicity of the four host stars is within 30\% of the metallicity of the Sun. The model also assumes that the exoplanets have undergone differentiation. 

Without going into the details of the model, which can be found in \citet{Valencia_2007}, it gives a proxy for the lowest amount of MMF possible for a given IMF. These constraints essentially suggest that a planet cannot have a very small or non-existent mantle. There is no natural process which can keep the water ice and core intact, while removing the mantle, a medium-density material, from an exoplanet. Similarly, the fixed Si$/$Fe ratio can be used as a proxy for the lowest possible values for MMF$/$CMF. For example, the constraints from Figure 1 of \citet{Valencia_2007} are MMF$/$IMF $>$ 0.2346 and MMF$/$CMF $>$ 0.5625. Note that for the extreme case of CMF$\rightarrow0$, these constrains imply MMF$\rightarrow0.19$ and IMF$\rightarrow0.81$.

Exoplanets in the habitable zone orbiting M-dwarf stars face strong stellar winds, mainly at the early stages of the planetary system. If the planet does not have a magnetic field of its own, the winds can vaporise volatiles like water from the exoplanet's surface \citep{Chassefiere2007, Lundin2007, Cohen_2020}. Collisions with celestial objects of large enough size can also add or remove mass to or from an already differentiated exoplanet. Nevertheless, the results we obtain here are valid even if the exoplanets' structure is altered significantly by impacts after differentiation or by erosion of its outer layers by stellar winds \citep{Valencia_2007}.

\subsubsection{Bayesian Inference}

Plotting exoplanets on mass radius curves only tells us if it is more or less likely to be made of lighter or denser material. Ternary diagrams, such as in Fig.\,\ref{fig:ternary}, show a range of possible compositions for the interior structure of exoplanets, given their observational mass, radius and uncertainties \citep{Zeng_2008, Valencia_2007, Rogers_2010}. Since more than one model parameter is uncertain and since different combinations of model parameters can give the same mass and radius of the exoplanet, n$\sigma$ bounds on ternary diagrams do not exactly represent a quantitative likelihood or probability of the composition. Bayesian inference considers this and provides a posterior likelihood for every parameter, namely CMF, MMF, and IMF and thus is a robust method for quantifying parameter degeneracy \citep{Rogers_2010, Dorn2015, Dorn2017}.
The Bayesian inference used in this study is based on \cite{Rogers_2010} although we use EOS and interior model from \citet{MagratheaHuang2022}. The likelihood function for this Bayesian analysis is given by
\begin{equation}
 {\scriptstyle L( \hat M_p, CMF, MMF) =} \frac {e^{- \frac{ ( M_p- \hat {M}_p )^{2}} {2 \sigma_{M_p}^{2}}- \frac { ( R_p- \hat {R}_p ( \hat {M}_p, CMF, MMF) )^{2}} {2 \sigma_{R_p}^{2}} }} {2 \pi \sigma _{M_p} \sigma _{R_p}}  { } ,
\end{equation}
in which the hats are used to represent the exoplanet mass as a free parameter and the calculated radius of the exoplanet. The Markov Chain Monte Carlo (MCMC) method with the Metropolis-Hastings algorithm is used for sampling from the posterior PDF. On each step the IMF is simply taken as $1-$CMF$-$MMF.  We first assess the interior structure of the exoplanets assuming no atmosphere. Later, we relax this assumption by considering the effect of adding an atmosphere of 1 or 2 percent mass fraction on the assessment of the interior of the exoplanets.

\subsection{Habitability}

In addition to the instellation of each planet (Table \ref{tab:table_properties}), we consider the contributions of tidal heating and atmospheric effects on the planets' surface temperature. We restrict ourselves to the biohabitable zone, which is a domain where liquid water can survive on at least part of the planetary surface \citep{Wandel_2018}. We do not consider complex geological or biological processes that affect habitability as they are beyond the scope of this paper.

\subsubsection{Tidal Heating}
The habitable zone around M-dwarf stars overlaps with the tidal locking radius. So, exoplanets around M-dwarf stars, which receive stellar radiation comparable to the Earth, are also affected by the differential gravity of the star. It produces heat at the expense of orbital eccentricity \citep{Henning_2009,Fabrycky2009}. 
We assess the effect of tidal heating on the equilibrium temperature of the exoplanets by using the tidal heating rates formulated by \citet{Peale_1978, Peale_1979,SHOWMAN_1997} and \citet{Henning_2009}. For a constant quality factor model it is given by 
\begin{equation}
 \dot {E}_{tidal} = \frac{21 k_{2} G M^{2}_{*} R^{5}_{p} n e^2} {2 Q a^6},
\end{equation}
where $Q$ is the tidal quality factor, $k_2$ is the second-order tidal love number, $M_{*}$ is the mass of the host star, $R_{p}$ is the radius of the exoplanet, $n$ is its mean motion, $e$ is its eccentricity and $a$ is its semi-major axis. We used a formalism based on \cite{Tobie_2019} to estimate $k_2$ and Q. 
Since the Q factor is sensitive to the assumed values of mantle viscosity ($\eta$) and the Andrade rheology parameter ($\alpha$), we consider our estimates as an order-of-magnitude estimate. When considering the effect of tidal heating on the exoplanets' total heat budget, we take $0.5$ and $5$ times the calculated values as their lower and upper bounds, respectively. Plate tectonic activity and degassing does impact habitability \citep{Sleep2001, Kasting2003, Driscoll2015, Unterborn_2022} but we do not consider them in this paper.  

\subsubsection{Atmospheric Impact} 

Potentially habitable exoplanets around M-dwarf stars may be in synchronous rotation, which can result in one side of the planet always facing the star with perpetual daylight. It also allows habitable conditions on parts of the exoplanet beyond the classical habitable zone \citep{Yang_2013, Wandel_2018}.
Tidally locked super-Earth exoplanet climate would drastically depend on the properties of the atmosphere. We use the analytical 1D model of \citet{Wandel_2018}, which calculates the surface temperature distribution using three factors: the irradiation from the host star (instellation), the atmospheric transmission (screening and greenhouse effect), and heat transport due to circulation and convection. The first factor is determined from the measured luminosity of the star and the planet's distance (Table \ref{tab:table_properties}) while the second and the third can in principle be calculated, given the planet's data (specific gravity, rotation) and the atmospheric properties (composition, pressure, heat capacity, wind speed, global circulation patterns, etc.). However, as these data are presently impossible to determine for exoplanets, they are represented by two parameters: atmospheric heating (or greenhouse factor) and the global heat transport factor ($f$).

The model gives an overall estimate of the temperature range on the planet, between the highest temperature at the substellar point and the lowest temperature at the night side. Of course, this simple model cannot calculate flow patterns like cells and vertical transport or more complex feedback mechanisms that depend on the composition, like clouds \citep{Yang_2013}. Sophisticated 3D Global Circulation Models (GCMs) may effectively do so \citep{Yang_2014,leconte2015, Kopparapu_2016, Kopparapu_2017, Turbet2018, Haqq-Misra_2018, Fauchez_2019, Wolf_2019}. Since we aim here at a comparative assessment of the four planets' habitability and not at an in-depth modelling of their atmospheres, we restrict ourselves to the simple model of \citet{Wandel_2018}. For completeness, we detail the main features of the model.

The analytic expression of the surface temperature is combined with the temperature boundaries of the HZ, to define a habitability range in the two-dimensional parameter plane, namely, atmospheric heating and circulation. \citet{Wandel_2018} defines the dimensionless heating factor $H$ which is a measure of the surface heating, combining the host star's irradiation with the albedo ($A$), the atmospheric screening ($\alpha$), and the greenhouse factor ($H_g$), as 
\begin{equation}
H = (1-A)\,H_g\,\alpha\,S / S_{\oplus} = H_{atm}\,s,
\end{equation}
where $S=L/4\pi r^2 $, and $s = S / S_{\oplus}$ is the instellation relative to Earth. The product $(1-A)H_g\alpha = H_{atm}$ is defined as the atmospheric albedo part of the heating factor. For comparison, some values of the heating factor for planets in the solar system are $H\sim1$ (Earth), $\sim0.3$ (Mars), and $\sim50$ (Venus). The surface temperature is calculated for each "latitude" (angular distance from the substellar point) by equating local heating and cooling. In the model, this is combined with the global heat transport, described by the parameter $f$, which gives the atmospheric circulation. It varies between $f=0$ (no heat transport) and $f=1$ (total transport giving an isothermal surface). While rocky planets with no or little atmosphere, like Mercury, have an extremely high day-night temperature contrast, planets with a thick, Venus-like atmosphere tend to be nearly isothermal. Intermediate cases, with up to $10$ bar atmospheres, conserve significant surface temperature gradients \citep{Selsis_2011}. The two extreme temperatures can be written as 
\begin{equation}
T_{min} = 278\,(Hf)^{1/4} {\rm \,K, \, \,and}
\end{equation}
\begin{equation}
T_{max} = 394\,H^{1/4} (1- \frac{3}{4} f) ^{1/4} {\rm \,K.}
\end{equation}

The range of temperatures allowing liquid water on at least part of the planet's surface could vary between freezing and boiling, or somewhat lower if the minimal moist greenhouse temperature ($\sim340$\,K) is taken \citep{Kopparapu_2013}. This temperature range defines the "biohabitability range" of the heating parameter. It extends between the lowest value, for which the substellar temperature is $273$\,K, and the highest value for which the substellar antipode is $373$\,K (or $\sim340$\,K for a more conservative range). In other words, the habitability range of the heating factor for locked planets may be defined as the range between the lower limit, for which the highest temperature (at the substellar point) is above the freezing point, and the upper limit of the heating factor, for which the lowest temperature (at the point opposite to the substellar point) is below boiling:
\begin{equation}
0.23\,(1 - \frac{3}{4} f)^{-1} < H < 3.2\,f^{-1}.
\end{equation}

\section{Results} \label{sec:results}

In what follows we detail the results we obtained, first for the possible internal structures of LHS\,1140\,b, K2-18\,b, TOI-1452\,b, and TOI-1468\,c, and then for their biohabitability.

\subsection{Internal Structure}

\begin{figure*}
  \centering
  \includegraphics[width=0.48\textwidth]{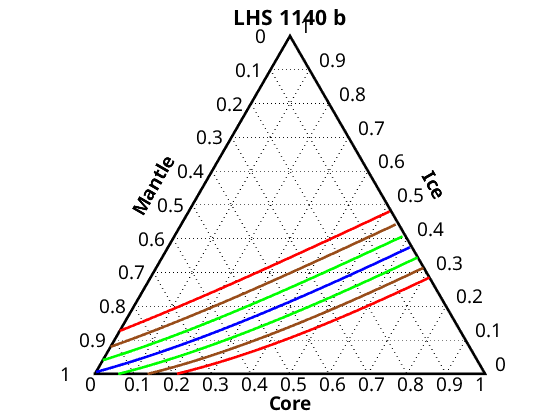}    
    \includegraphics[width=0.48\textwidth]{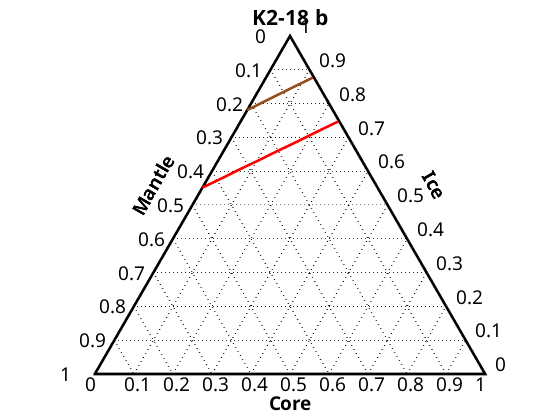}
    \\
    \includegraphics[width=0.48\textwidth]{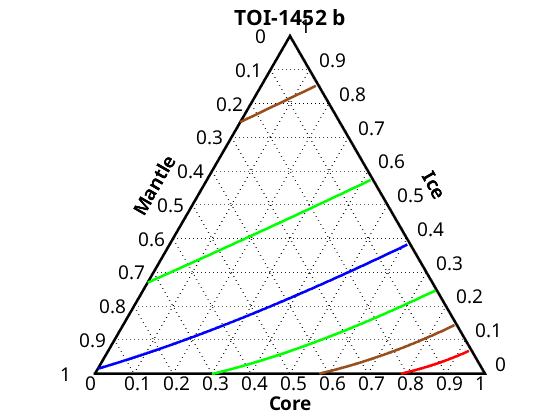}
    \includegraphics[width=0.48\textwidth]{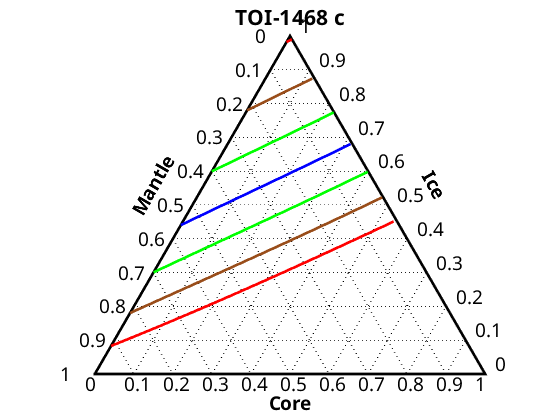}
    \caption{Ternary diagrams for LHS\,1140\,b, K2-18\,b, TOI-1452\,b, and TOI-1468\,c based on a code by \cite{Zeng_2008}. Each point on the ternary diagram represents a unique composition. Horizontal, positive-slope, and negative-slope lines describe the planet's water-ice, iron, and mantle mass fractions, respectively. The blue curve shows compositions that explain the observed mass and radius of the exoplanet. Green, brown, and red curves represent compositions for one, two, and three standard deviations in observed mass and radius.}
    \label{fig:ternary}
        
\end{figure*}

We first calculated ternary diagrams of the exoplanets to visualise the possible compositions that explain the observed mass and radius considering their uncertainties. Figure \ref{fig:ternary} shows the ternary diagrams for LHS\,1140\,b, K2-18\,b, TOI-1452\,b, and TOI-1468\,c based on a publicly available code by \cite{Zeng_2008}. The interior structure model and EOS for obtaining the ternary diagrams are based on \citet{MagratheaHuang2022} as mentioned in Section \ref{sec:models}. The diagrams are read as follows: IMF increases from 0 to 1 from the base to the apex of the triangle, and each point on the ternary diagram represents a unique composition. For example, the point labelled 0.2 on the base of the ternary diagram of LHS\,1140\,b represents one single composition consisting of 0.0 IMF, 0.8 MMF, and 0.2 CMF. The blue curve shows compositions that explain the observed mass and radius of the exoplanet. It is a curve and not a point because of the degeneracy in the composition of each exoplanet. For example, the observed mass and radius without standard deviation of LHS\,1140\,b can be explained by IMF spanning from 0 to 0.38, MMF from 0 to 1.0, and CMF from 0.0 to 0.62. Green, brown, and red curves represent compositions for one, two, and three standard deviations in observed mass and radius. Large uncertainties in the observed mass and radius result in a large spread of the standard deviation curves.

Since K2-18\,b has a low bulk density ($2.67 g/cm^3$) comparable to a high-pressure polymorph of ice, the blue curve representing its composition lies outside the ternary diagram. Its low-density standard deviation curves also lie beyond 100 percent IMF and hence are not shown in the ternary diagram. A significant amount of atmosphere is needed to explain those lower densities. We explore the possibility of atmosphere in the subsequent sections. Similarly, the ternary diagrams of TOI-1452\,b and TOI-1468\,c, show that the density uncertainties of these exoplanets result in a wide range of possible compositions.


\begin{figure*}
    \centering
    \includegraphics[width=0.48\textwidth]{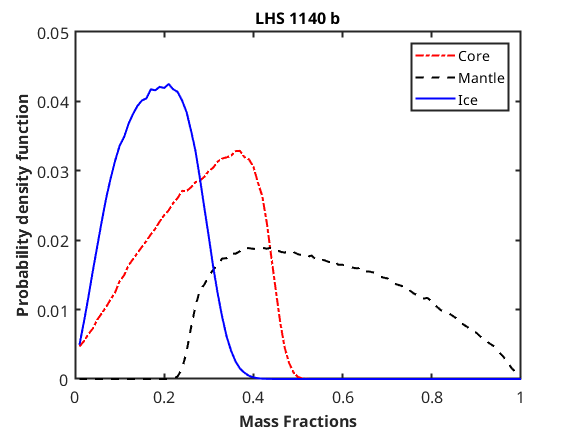}    
    \includegraphics[width=0.48\textwidth]{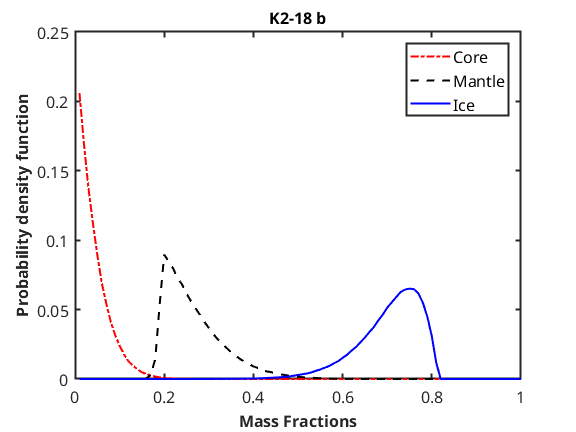}
    \\
    \includegraphics[width=0.48\textwidth]{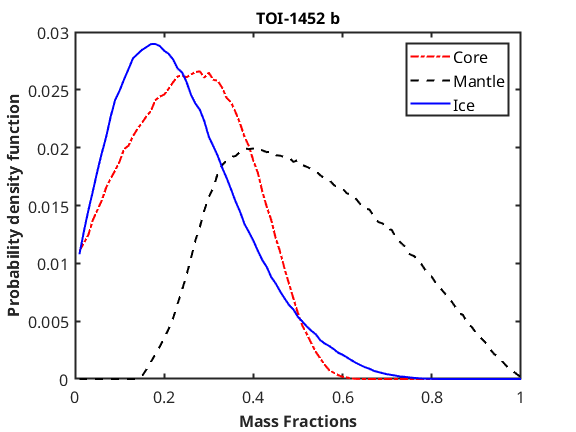}
    \includegraphics[width=0.48\textwidth]{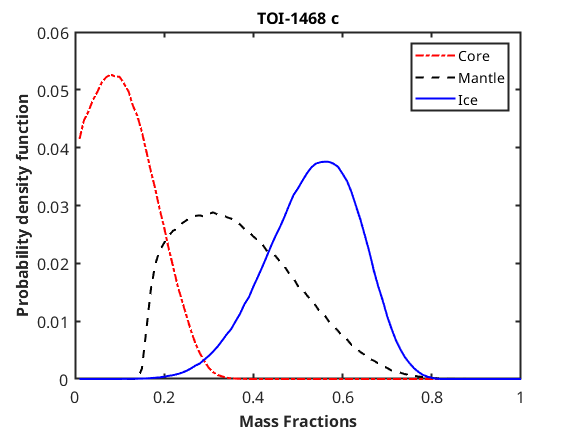}
    \caption{Posterior probability density functions of core mass fraction (red dash-dot curve), mantle mass fraction (black dash curve) and water-ice mass fraction (blue solid curve) of exoplanets LHS\,1140\,b (upper left), K2-18\,b (upper right), TOI-1452\,b (lower left) and TOI-1468\,c (lower right), considering no atmosphere.}
    \label{fig:mass_fractions}
\end{figure*}

The interior structure model and EOS in our Bayesian analysis are based on \citet{MagratheaHuang2022} as mentioned in Section \ref{sec:models}. Figure \ref{fig:mass_fractions} shows, for each of the four planets, the posterior probability density functions (PDFs) of mass fractions of the core (red dash-dot curve), mantle (black dash curve), and water-ice (blue solid curve) layers, considering constraints based on relative elemental abundance and considering no contribution of the atmosphere to the observed masses and sizes of the exoplanets. 
The binning interval for sampling is 1 percent. 
For LHS\,1140\,b, as an example, there is $\sim0.5$\% probability that the planet will have an IMF $<1$\%. This is consistent with the composition curves touching the base of the ternary diagram in Figure \ref{fig:ternary}, where the IMF is 0. It means that there is a small but finite probability that the exoplanet can be completely rocky with virtually no water on its surface. 
For comparison, the low density of K2-18\,b results in a relatively small core and an extremely large IMF within the scope of our modelling. Unlike for the other three planets, the marginal IMF, CMF, and MMF PDFs of K2-18\,b are highly skewed, being pushed to their allowed limits from the relative elemental abundance constraints we used. This may indicate that the simplistic 1D modelling we use here may be insufficient for the extremely low-density planets such as K2-18\,b, and we discuss it in Section 4.

For TOI-1452\,b, and TOI-1468\,c, the peak heights of the PDFs are low compared to that of LHS\,1140\,b because uncertainties in their observations are higher and hence the compositions are not well constrained. Similarly to LHS\,1140\,b, there is a non-zero probability that TOI-1452\,b will have an IMF $<1$\,\%. The IMF PDF peaks below $25$\,\% for both planets. This means that both LHS\,1140\,b and TOI-1452\,b are likely to be super-Earths instead of mini-Neptunes. On the other hand, our Bayesian inference, considering no atmosphere, excludes the possibility that K2-18\,b and TOI-1468\,c are rocky worlds, because there is virtually zero probability that their IMF is between 0\% and 1\%. In general, the parts of the graph for core, mantle and ice, which have 0 probability show that Bayesian modelling of the interior of an exoplanet, also using constraints based on planet formation conditions, can be useful to rule out some compositions of an exoplanet.


\begin{figure*}
    \centering
    \includegraphics[width=0.48\textwidth]{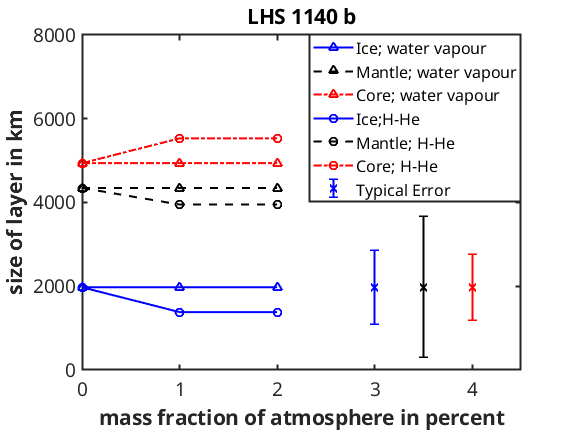}    
    \includegraphics[width=0.48\textwidth]{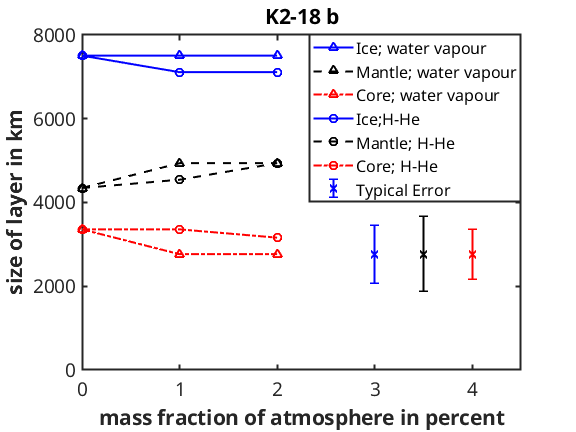}
    \\
    \includegraphics[width=0.48\textwidth]{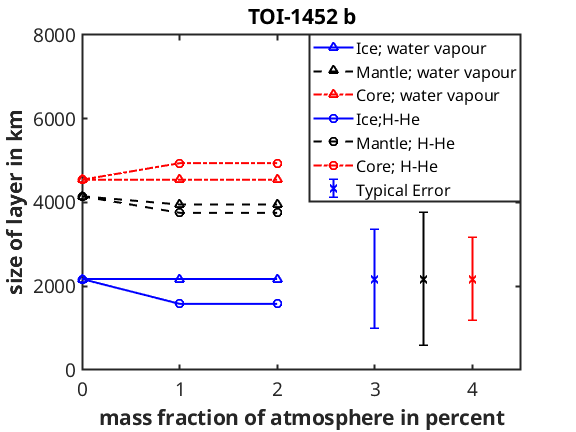}
    \includegraphics[width=0.48\textwidth]{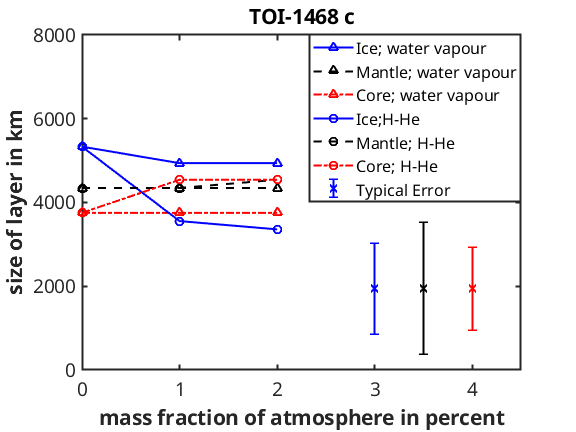}
    \caption{Median sizes of layers of LHS\,1140\,b (upper left), K2-18\,b (upper right), TOI-1452\,b (lower left) and TOI-1468\,c (lower right) as a function of the atmosphere mass fraction. Triangles and circles represent the sizes of layers considering a water vapor and a H-He atmosphere, respectively. Typical error bars are shown on the right.}
    \label{fig:atmosphere}
\end{figure*}

In addition to the their mass-fraction PDFs, we calculated the posterior PDFs of the thickness of the core, mantle and water-ice layers in the no-atmosphere scenario (see Figure \ref{fig:atmosphere} and Table \ref{tab:sizes_vs_AMF}). Furthermore, to relax the no-atmosphere assumption, we explored the interior structure of these exoplanets considering an atmosphere of $1$\,\% or $2$\,\% in mass fraction. While the inclusion of an atmosphere does not change the mass of the planet, it slightly reduces the radius of its solid+liquid phases, compared with the transit radius, which in turn may affect the range of possibilities for the internal structure.

We have considered the atmosphere as the fourth and the outermost layer. We used MAGRATHEA to model the atmosphere.
It enables the calculation of the contribution of the atmospheric layer in the transit radius. For simplicity, we only consider two possible atmospheres: H-He and water vapor. While the H-He atmosphere represents a low molecular weight and less conductive ($f$) atmosphere, the water vapor atmosphere represents a high molecular weight and conductive atmosphere, such as CO$_2$, H$_2$O or CH$_4$ rich atmospheres would be.

\label{appendix:cg}

In Figure \ref{fig:atmosphere} we plot the most likely sizes of layers as a function of the mass fraction of the atmosphere considered. Estimates of mass fractions and sizes of layers only change slightly after considering an atmosphere. 
Across the four exoplanets, adding an atmosphere decreases the most likely ice mass fraction and the size of that layer, particularly when considering a H-He atmosphere. With the exception of K2-18\,b, the mass fraction and the size of the core increases when the mass fraction of the atmosphere is increased from zero. Since a H-He atmosphere has a lower density than a water vapor atmosphere, the effect of a H-He atmosphere is larger. As expected, a higher mass fraction of the atmosphere results in a smaller ice layer and a larger core to keep the bulk density constant. The size of the mantle does not change as much as the ice layer or core, probably because of its medium density. Interestingly, for all four exoplanets, except for K2-18\,b, an increase in the amount of a H-He atmosphere increases the probability of the IMF being $<1$\,\%, thus increasing the possibility of these planets having a rocky surface. Particularly for LHS\,1140\,b and TOI-1452\,b, the presence of a 1\% or 2\% H-He atmosphere reduces the size of the ice layer and increases the probability that these exoplanets have a rocky surface to more than 3\%.

Table \ref{tab:sizes_vs_AMF} lists the most likely sizes of the three inner layers and their uncertainties, for all four exoplanets and for the different atmospheres we have considered thus far. In addition to atmospheres of $1$\,\% and $2$\,\% in mass fraction, we considered $0.1$\,\% and $10$\,\% mass-fraction atmospheres, but found very similar results. The effect of a $0.1$\,\% atmosphere on the median sizes and mass fractions of the interior layers was the same as a $1$\,\% atmosphere, and the effect of a $10$\,\% atmosphere was similar to the effect of a $2$\,\% atmosphere. The results for $0.1$\,\% and $10$\,\% atmospheres are excluded from Figure \ref{fig:atmosphere} and Table \ref{tab:sizes_vs_AMF} for brevity.

\begin{table*}
    \centering
    \caption{Sizes of the three inner layers in thousand kilometers considering different atmospheres}
    \begin{tabular}{l c ccccc c}
        \hline
        & Planets & LHS\,1140\,b & K2-18\,b & TOI-1452\,b & TOI-1468\,c & \\
        \hline
        Atmosphere & Layer & Thickness [km] & Thickness [km] & Thickness [km] & Thickness [km] & \\
        \hline
        \multicolumn{1}{l}{No atmosphere} & Ice & $2.0^{+1.0}_{-0.8}$ & $7.5^{+0.6}_{-0.8}$ & $2.2^{+1.4}_{-1.0}$ & $5.3^{+1.0}_{-1.2}$ & \\
         & Mantle & $4.3^{+2.0}_{-1.4}$ & $4.3^{+1.0}_{-0.8}$ & $3.1^{+1.8}_{-1.4}$ & $4.3^{+1.8}_{-1.4}$ & \\
         & Core & $4.9^{+0.6}_{-1.0}$ & $3.4^{+0.6}_{-0.6}$ & $4.5^{+0.8}_{-1.2}$ & $3.8^{+0.8}_{-1.2}$ & \\
        \hline
        1\% Water Vapor & Ice & $1.0^{+0.6}_{-0.7}$ & $7.5^{+0.8}_{-1.0}$ & $2.2^{+1.2}_{-1.2}$ & $4.9^{+1.2}_{-1.0}$ & \\
         & Mantle & $3.6^{+1.2}_{-0.6}$ & $4.9^{+1.8}_{-1.0}$ & $4.0^{+1.8}_{-1.2}$ & $4.3^{+2.0}_{-1.4}$ & \\
         & Core & $5.0^{+0.6}_{-1.0}$ & $2.8^{+1.0}_{-0.8}$ & $4.5^{+0.8}_{-1.2}$ & $3.8^{+1.0}_{-1.2}$ & \\
        \hline
        2\% Water Vapor & Ice & $2.0^{+0.8}_{-1.0}$ & $7.5^{+0.8}_{-1.0}$ & $2.2^{+1.2}_{-1.2}$ & $4.9^{+1.2}_{-1.0}$ & \\
         & Mantle & $4.3^{+1.8}_{-1.4}$ & $4.9^{+1.8}_{-1.0}$ & $4.0^{+1.8}_{-1.2}$ & $4.3^{+2.0}_{-1.4}$ & \\
         & Core & $4.9^{+0.6}_{-1.0}$ & $2.8^{+1.0}_{-0.8}$ & $4.5^{+0.8}_{-1.2}$ & $3.8^{+0.8}_{-1.2}$ & \\
        \hline
        1\% H-He & Ice & $1.4^{+0.6}_{-0.8}$ & $7.1^{+0.8}_{-1.2}$ & $1.6^{+1.0}_{-1.0}$ & $3.6^{+1.0}_{-1.2}$ & \\
         & Mantle & $4.0^{+1.4}_{-0.3}$ & $4.5^{+1.2}_{-1.0}$ & $3.8^{+1.4}_{-1.0}$ & $4.3^{+2.2}_{-1.6}$ & \\
         & Core & $5.5^{+0.4}_{-0.8}$ & $3.4^{+0.8}_{-0.8}$ & $4.9^{+0.8}_{-1.2}$ & $4.5^{+0.8}_{-1.2}$ & \\
        \hline
        2\% H-He & Ice & $1.4^{+0.6}_{-0.8}$ & $7.1^{+0.8}_{-1.2}$ & $1.6^{+1.0}_{-1.0}$ & $3.4^{+1.2}_{-1.2}$ & \\
         & Mantle & $4.0^{+1.4}_{-1.0}$ & $4.9^{+1.6}_{-1.4}$ & $3.8^{+1.4}_{-1.0}$ & $4.5^{+0.8}_{-1.2}$ & \\
         & Core & $5.5^{+0.2}_{-0.8}$ & $3.2^{+1.0}_{-1.2}$ & $4.9^{+0.8}_{-1.2}$ & $4.5^{+0.8}_{-1.2}$ & \\
        \hline
    \end{tabular}
    \label{tab:sizes_vs_AMF}
\end{table*}



\subsection{Habitability}

All four exoplanets fall within the tidal locking zone and the habitable zone \citep{Wandel_2023Extended},
hence  tidal heating may influence the global heat budget and habitability. Tidal heating strongly depends on eccentricity. Since for TOI-1452\,b \citep{Cadieux_2022} and TOI-1468\,c \citep{chaturvedi_2022} the authors only considered circular orbits to fit observational data, we considered the more physical eccentricity of $0.01$ for tidal heat calculations. Similarly, we took the upper limit of $0.043$ as the eccentricity of LHS\,1140\,b in calculating its tidal heating. For K2-18\,b we used the measured value of $e=0.2$. Table \ref{tab:tidal_heating} shows the calculated values of the second order tidal love number ($k_2$), Quality factor (Q), tidal heating rate ($\dot E_{tidal}$), 
and instellation received from the host star ($\dot E_{ins}$). The Table also shows the global mean temperature (GMT) of the planets, after considering tidal heating, instellation, and $0.3$ albedo, but not any other atmospheric effects.  For all four exoplanets, compared with the instellation, the tidal heating is very small and does not significantly affect the surface temperature. Only for K2-18\,b, considering 5 times the tidal heating calculated here as an upper limit, there is asignificant temperature increase, of $+2$\,K.


\begin{table*}
    \centering
    \caption{Stellar instellation and tidal heating}
    \begin{tabular}{ccccccc}
         \hline
         Exoplanets &  $k_2$ &  Q factor & $\dot{E}_{tidal}$ [W] & $\dot{E}_{ins}$ [W] & GMT$^*$ [K] \\
         \hline
         LHS\,1140\,b & 0.3688 & 131 & $2.9044 \times 10^{13}$ & $1.6255 \times 10^{17}$ & 208   \\
         K2-18\,b & 0.0830& 8 & $8.2883 \times 10^{15}$& $1.4718 \times 10^{18}$ & 270 \\
         TOI-1452\,b & 0.3723 & 170  & $5.8473 \times 10^{13}$ & $9.1709 \times 10^{17}$ & 298 \\
         TOI-1468\,c & 0.2268 & 11  & $2.6837 \times 10^{14}$ & $1.6058 \times 10^{18}$ & 309 \\
         \hline
    \end{tabular}\\
    Notes. $^*$ Considering instellation, an albedo of 0.3, and tidal heating, but not the contribution from atmospheric effects.
    \label{tab:tidal_heating}
\end{table*}

 

\begin{figure}
\includegraphics[width=1\linewidth]{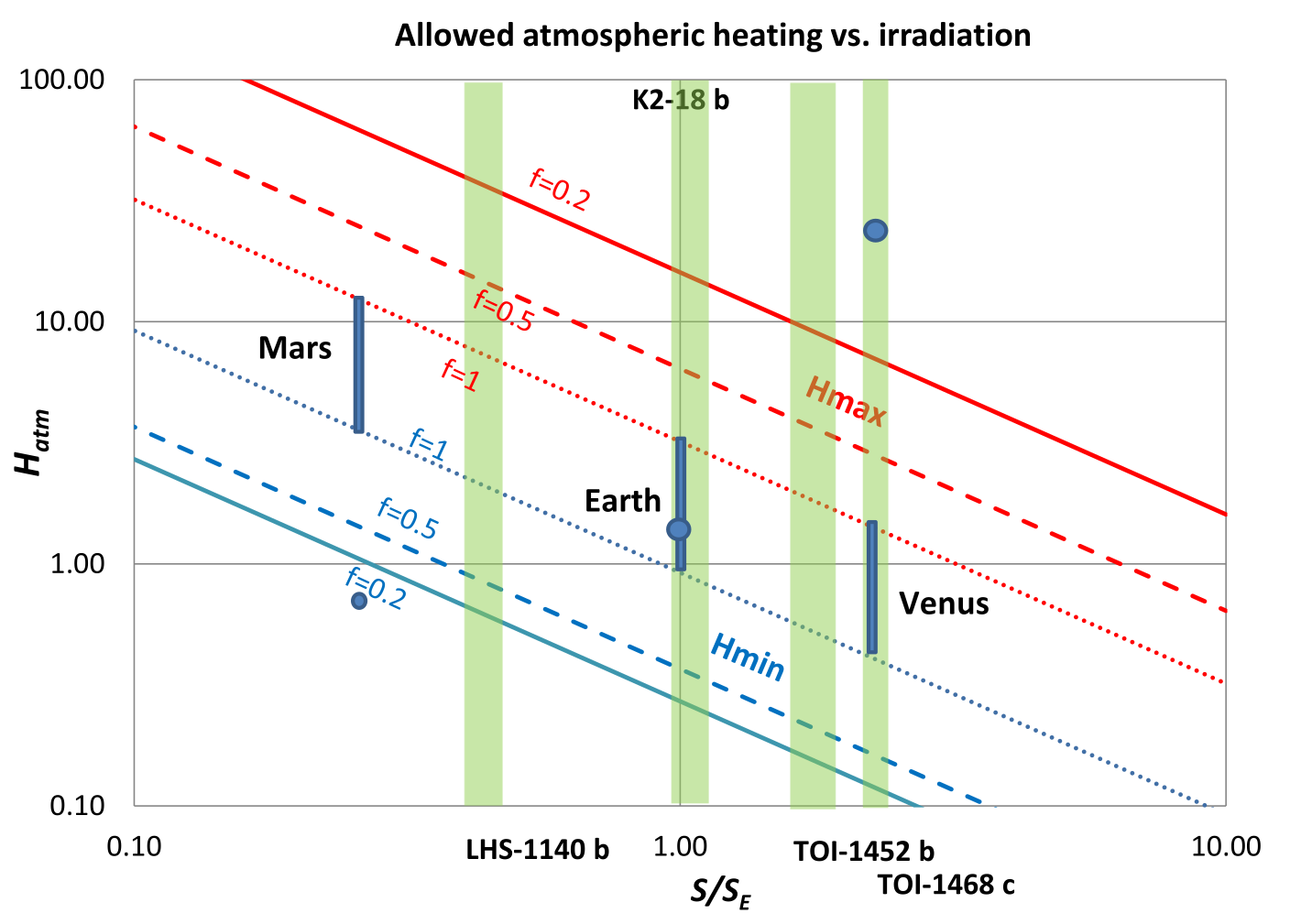}
\caption{Maximal (red) and minimal (blue) atmospheric heating factor vs instellation, for three values of the heat redistribution parameter: $f$ = 0.2 (solid), 0.5 (dashed), and 1 (dotted). The locations of the four planets in Table \ref{tab:table_properties} considering their instellation are marked by vertical green stripes. Modified from Fig. 11 of \citet{Wandel_2018}. The allowed habitable $H_{atm}$ ranges of Earth, Mars and Venus for $f$ = 1 are marked by vertical blue strips. The effective values of $H_{atm}$ for these terrestrial planets are represented by blue dots. \label{fig:se2}}
\end{figure}

Figure \ref{fig:se2} shows the range of atmospheric heating vs instellation, which will enable biohabitable conditions on at least some parts of a tidally-locked planet for different values of the heat redistribution parameter ($f$). Maximal and minimal atmospheric heating factors are represented by red and blue lines. Ranges for three possible $f$ values are shown: $f = 0.2, 0.5$, and $1$ in solid, dashed, and dotted lines, respectively. The locations of the four planets in Table \ref{tab:table_properties} are marked by vertical green stripes. The width of each stripe represents the uncertainty in its instellation measurement. For reference, Earth, Mars and Venus are represented in the graph as vertical blue lines between the maximal and minimal atmospheric heating factor of 1. Even though Earth and Mars are not tidally-locked, we find their addition useful for better understanding the figure. The effective values of $f$ for Earth, Venus, and Mars are estimated as 0.95, 1, and 0.85, respectively, and their respective values of atmospheric heating factor ($H_{atm}$) are 1.15, 26, and 0.75 \citep{Wandel_2018}. Their actual $H_{atm}$ values are represented by blue dots in Figure \ref{fig:se2}.

The figure shows that tidally locked exoplanets have a wider instellation range, for which biohabitable conditions may exist somewhere on the planet. Planets orbiting M-dwarf stars may have temperatures compatible with liquid water, at least on part of their surface, for a wide range of atmospheric properties compared to planets that are not tidally locked. Because of rapid rotation and global circulation $f$ is typically close to 1 for planets that are not tidally locked, whereas it can be significantly less than 1 for tidally locked planets, especially for thin atmospheres. For tidally locked planets, if the instellation is high enough to make the substellar point uninhabitable, it might still be habitable on the night side, if $f<1$. Conversely, if the instellation is so low that the permanent night side is too cold to be habitable, the substellar point could be habitable if $f<1$, a phenomenon recently named "eyeball planets" \citep{eyeball_planet2013}.


\section{Summary and Discussion} \label{sec:discussion}

We have assessed the interior structure and biohabitability of the super-Earth exoplanets LHS\,1140\,b, K2-18\,b, TOI-1452\,b, and TOI-1468\,c. Their interior structure was estimated by using Bayesian inference, taking into account constraints based on relative elemental abundance. Assuming no atmosphere or thin atmospheres, K2-18\,b and TOI-1468\,c are likely water worlds. 
In contrast with these two, there is a finite (albeit small) probability that TOI-1452\,b and LHS\,1140\,b are terrestrial exoplanets with a rocky surface.

LHS\,1140 and K2-18 have slightly subsolar and supersolar metallicities, respectively. Since we assumed solar relative elemental abundances for our calculations, the actual core size and mass fraction of LHS\,1140\,b are likely somewhat smaller than we estimated, and those for K2-18\,b are likely somewhat larger. Figure \ref{fig:mass_fractions} indicates that for K2-18\,b the solution with CMF\,$\rightarrow0$ is the most likely one. Given the constrains we used from relative elemental abundances, this implies MMF\,$\rightarrow0.19$ and IMF\,$\rightarrow0.81$. Although coreless terrestrial planets were discussed as a possible outcome of planet formation and evolution processes \citep{2008ApJ...688..628E}, the pebble accretion model for
terrestrial planet formation suggests that the CMF should increase with the total mass of the planet \citep{2023A&A...671A..74J}, which makes the CMF\,$=0$ solution unlikely for the four planets we analyzed here. We conclude that the simplistic 1D modelling we used here may be insufficient to infer the internal structure of extremely low-density planets such as K2-18\,b. Its highly skewed marginal PDFs in Figure \ref{fig:mass_fractions}, pushed to their allowed limits from relative elemental abundance constraints, possibly indicate that a massive atmosphere is needed to explain its observed mass and radius \citep[e.g.,][]{Madhusudhan_2023}.


The so-called radius valley is a scarcity in the number of exoplanets between 1.5 and 2.0 Earth’s radii \citep{Fulton_2017}. LHS\,1140\,b and TOI-1452\,b lie in the middle of the valley. For exoplanets of a radius of more than two Earth radii, like K2-18\,b and TOI-1468\,c, it is possible that they have retained a thick, mostly H-He, primary atmosphere. \citet{Rogers2015Radii} claims that exoplanets with a radius of more than $1.6$ Earth radii are not entirely rocky, and that they should have at least a small amount of volatiles, like water and atmospheric gases, in their envelope. It is therefore interesting to investigate how LHS\,1140\,b and TOI-1452\,b got their atmospheric mass fractions and water mass fractions just right to have their transit radii in the middle of the radius valley. For instance, \citet{Luque_2022} proposed that it is actually a density gap that separates rocky from water-rich exoplanets. 

As for their biohabitability, we can calculate the parameters $f$ and $H_{atm}$ for these exoplanets only if their atmospheric mass and composition is known. Nevertheless, comparing with the values of these parameters for planets in the solar system we can draw some conclusions about these exoplanets. Considering the escape velocity of the exoplanets and their equilibrium temperature, all of them are capable of retaining H-He in their atmospheres \citep{Bourrier2017, Konatham_2020}. However, its actual presence is likely to depend on the presence of a magnetic field and the evolutionary trajectory of the exoplanet. Since these exoplanets are massive, their greenhouse factor ($H_g$) is likely to be more than $1$. A minimum value of $H_g\sim 1$ will be for the case of a pure H-He atmosphere. Water worlds could also have a high transport factors ($f$) because of atmospheric and ocean circulation, so the temperature distribution is expected to be relatively homogeneous over the planet.

Assuming that the four planets studied here retained thick atmospheres, they may have highly transporting atmospheres ($f\simeq1$) with significant greenhouse factors ($H_g>1$). 
Combined with their higher instellation, Figure \ref{fig:se2} shows that for a sufficiently high $H_g$ and highly transporting atmospheres, such as can be the case for water or CO$_{2}$ rich atmospheres, K2-18\,b, TOI-1468\,c, and TOI-1452\,b would be too hot to sustain liquid water on their surface. However, for the case of an almost pure H-He atmosphere a lower value of $H_g\simeq1$ is possible, which in principle should enable liquid water on some parts of the planets. Indeed, based on \textit{JWST} observations, \citet{Madhusudhan_2023} show that there is a limited parameter range over which a habitable ocean could still exist on K2-18\,b. On the other hand, the lower instellation of LHS\,1140\,b and its finite probability of having a rocky surface give more space for biohabitable conditions in the case of $f\sim1$ (see Figure \ref{fig:se2}). Although \textit{HST} observations found marginal evidence for water vapor in its atmosphere \citep{Edwards_2021}, the analysis of \textit{JWST} observations could only place a $2\sigma$ upper limit on logH$_2$O of $-2.94$ \citep{Cadieux_JWST_2024}. The observations ruled out a H$_2$ dominated atmosphere in favour of a N$_2$ dominated one. However, they do not rule out the existence of atmospheric CO$_2$ or H$_2$O and even liquid-water oceans \citep{Damiano_2024}.



Being nearby, the transiting planets we have analyzed here are potential targets for in-depth exoplanetary characterization. Future spectroscopic studies of these exoplanets might provide better insights into their atmospheric conditions. The \textit{JWST} and the upcoming ELT are designed to detect molecular species in the atmospheres of these exoplanets \citep{Lillo-Box_2020, Wunderlich_2019}. Of the four planets studied here, we find LHS\,1140\,b an excellent target for future searches of biosignatures. Relating the atmospheric properties of the planets with their interior structures can be useful in making the knowledge of the interior and surface conditions more accurate. Similarly, precise phase-curve measurements can be used to estimate the actual temperature difference between the planet's permanent day and permanent night sides \citep{Rauscher_2008, Hammond_2017}.

In the future, measurements of radio emission caused by maser instabilities in the atmospheres of exoplanets will help to confirm the existence of magnetic fields generated by their cores \citep{Lynch2017, Varela2018, Lazio2019, Turner_2021}. This is expected to become a milestone in exoplanets' habitability and interior structure studies, since the existence of a magnetic field is considered to help a planet retain its atmosphere and the strength of the magnetic field may enable it to constrain the size of its core. By extending the timespan and improving the precision of radial velocity or transit data, the interior structure of the planets can be further constrained by using the second-order tidal love number (k$_2$) to find an upper bound on their CMF \citep{Kramm_2011, Bernabo2024}.

\begin{acknowledgments}
\textit{Acknowledgments}. We are grateful to Dr. Joe P Renaud for his insightful feedback and guidance in assessing tidal heat. This research was partially supported by the Israel Science Foundation through grant No. 1404/22.
Based on observations made with ESO Telescopes at the La Silla Paranal Observatory under programme IDs 191.C-0873, 0100.C-0884, 198.C-0838, 60.A-9709. VP acknowledges funding via the Kimmel Prize Fellowship and the Fellowship of the Dean of Chemistry at the Weizmann Institute of Science. Some of the data presented herein were obtained at the W. M. Keck Observatory, which is operated as a scientific partnership among the California Institute of Technology, the University of California and the National Aeronautics and Space Administration. The Observatory was made possible by the generous financial support of the W. M. Keck Foundation. 
A. Wandel's research was supported by the Minerva Grant to the Center of Planetary Origin of Life at the Hebrew University of Jerusalem.

\end{acknowledgments}

\bibliography{bibiography}{}

\begin{thebibliography}{}
\expandafter\ifx\csname natexlab\endcsname\relax\def\natexlab#1{#1}\fi
\providecommand{\url}[1]{\href{#1}{#1}}
\providecommand{\dodoi}[1]{doi:~\href{http://doi.org/#1}{\nolinkurl{#1}}}
\providecommand{\doeprint}[1]{\href{http://ascl.net/#1}{\nolinkurl{http://ascl.net/#1}}}
\providecommand{\doarXiv}[1]{\href{https://arxiv.org/abs/#1}{\nolinkurl{https://arxiv.org/abs/#1}}}

\bibitem[{{Acuña, Lorena} {et~al.}(2021){Acuña, Lorena}, {Deleuil, Magali},
  {Mousis, Olivier}, {Marcq, Emmanuel}, {Levesque, Maëva}, \& {Aguichine,
  Artyom}}]{Acuna2021}
{Acuña, Lorena}, {Deleuil, Magali}, {Mousis, Olivier}, {et~al.} 2021, A\&A,
  647, A53, \dodoi{10.1051/0004-6361/202039885}

\bibitem[{{Adibekyan} {et~al.}(2015){Adibekyan}, {Santos}, {Figueira}, {Dorn},
  {Sousa}, {Delgado-Mena}, {Israelian}, {Hakobyan}, \&
  {Mordasini}}]{Adibekyan2015}
{Adibekyan}, V., {Santos}, N.~C., {Figueira}, P., {et~al.} 2015, \aap, 581, L2,
  \dodoi{10.1051/0004-6361/201527059}

\bibitem[{{Alibert, Y.}(2014)}]{Alibert_2014}
{Alibert, Y.} 2014, A\&A, 561, A41, \dodoi{10.1051/0004-6361/201322293}

\bibitem[{{Anderson} {et~al.}(2004){Anderson}, {Schubert}, {Jacobson}, {Lau},
  {Moore}, \& {Palguta}}]{2004Sci...305..989A}
{Anderson}, J.~D., {Schubert}, G., {Jacobson}, R.~A., {et~al.} 2004, Science,
  305, 989, \dodoi{10.1126/science.1099050}

\bibitem[{{Anderson} {et~al.}(1998){Anderson}, {Schubert}, {Jacobson}, {Lau},
  {Moore}, \& {Sjogren}}]{1998Sci...281.2019A}
---. 1998, Science, 281, 2019, \dodoi{10.1126/science.281.5385.2019}

\bibitem[{{Angerhausen} {et~al.}(2013){Angerhausen}, {Sapers}, {Citron},
  {Bergantini}, {Lutz}, {Queiroz}, {Alexandre}, \&
  {Araujo}}]{eyeball_planet2013}
{Angerhausen}, D., {Sapers}, H., {Citron}, R., {et~al.} 2013, Astrobiology, 13,
  309, \dodoi{10.1089/ast.2012.0846}

\bibitem[{{Barnes}(2017)}]{2017CeMDA.129..509B}
{Barnes}, R. 2017, Celestial Mechanics and Dynamical Astronomy, 129, 509,
  \dodoi{10.1007/s10569-017-9783-7}

\bibitem[{{Benneke} {et~al.}(2019){Benneke}, {Wong}, {Piaulet}, {Knutson},
  {Lothringer}, {Morley}, {Crossfield}, {Gao}, {Greene}, {Dressing},
  {Dragomir}, {Howard}, {McCullough}, {Kempton}, {Fortney}, \&
  {Fraine}}]{Benneke_2019}
{Benneke}, B., {Wong}, I., {Piaulet}, C., {et~al.} 2019, \apjl, 887, L14,
  \dodoi{10.3847/2041-8213/ab59dc}

\bibitem[{{Bernab{\`o}} {et~al.}(2024){Bernab{\`o}}, {Csizmadia}, {Smith},
  {Rauer}, {Hatzes}, {Esposito}, {Gandolfi}, \& {Cabrera}}]{Bernabo2024}
{Bernab{\`o}}, L.~M., {Csizmadia}, S., {Smith}, A.~M.~S., {et~al.} 2024, \aap,
  684, A78, \dodoi{10.1051/0004-6361/202346852}

\bibitem[{Bezacier {et~al.}(2014)Bezacier, Journaux, Perrillat, Cardon,
  Hanfland, \& Daniel}]{Bezacier2014}
Bezacier, L., Journaux, B., Perrillat, J.-P., {et~al.} 2014, The Journal of
  Chemical Physics, 141, 104505, \dodoi{10.1063/1.4894421}

\bibitem[{Boujibar {et~al.}(2020)Boujibar, Driscoll, \& Fei}]{Boujibar2020}
Boujibar, A., Driscoll, P., \& Fei, Y. 2020, Journal of Geophysical Research:
  Planets, 125, e2019JE006124, \dodoi{https://doi.org/10.1029/2019JE006124}

\bibitem[{{Bourrier} {et~al.}(2017){Bourrier}, {Ehrenreich}, {King},
  {Lecavelier des Etangs}, {Wheatley}, {Vidal-Madjar}, {Pepe}, \&
  {Udry}}]{Bourrier2017}
{Bourrier}, V., {Ehrenreich}, D., {King}, G., {et~al.} 2017, \aap, 597, A26,
  \dodoi{10.1051/0004-6361/201629253}

\bibitem[{Cadieux {et~al.}(2022)Cadieux, Doyon, Plotnykov, Hébrard, Jahandar,
  Étienne Artigau, Valencia, Cook, Martioli, Vandal, Donati, Cloutier, Narita,
  Fukui, Hirano, Bouchy, Cowan, Gonzales, Ciardi, Stassun, Arnold, Benneke,
  Boisse, Bonfils, Carmona, Cortés-Zuleta, Delfosse, Forveille, Fouqué,
  da~Silva, Jenkins, Kiefer, Ágnes Kóspál, Lafrenière, Martins, Moutou,
  do~Nascimento, Ould-Elhkim, Pelletier, Twicken, Bouma, Cartwright,
  Darveau-Bernier, Grankin, Ikoma, Kagetani, Kawauchi, Kodama, Kotani, Latham,
  Menou, Ricker, Seager, Tamura, Vanderspek, \& Watanabe}]{Cadieux_2022}
Cadieux, C., Doyon, R., Plotnykov, M., {et~al.} 2022, The Astronomical Journal,
  164, 96, \dodoi{10.3847/1538-3881/ac7cea}

\bibitem[{Cadieux {et~al.}(2024{\natexlab{a}})Cadieux, Plotnykov, Doyon,
  Valencia, Jahandar, Dang, Turbet, Fauchez, Cloutier, Cherubim, Étienne
  Artigau, Cook, Edwards, Hallatt, Charnay, Bouchy, Allart, Mignon, Baron,
  Barros, Benneke, Martins, Cowan, Medeiros, Delfosse, Delgado-Mena, Dumusque,
  Ehrenreich, Frensch, Hernández, Hara, Lafrenière, Curto, Malo, Melo,
  Mounzer, Passeger, Pepe, Poulin-Girard, Santos, Sosnowska, Mascareño,
  Thibault, Vaulato, Wade, \& Wildi}]{Cadieux_2024}
Cadieux, C., Plotnykov, M., Doyon, R., {et~al.} 2024{\natexlab{a}}, The
  Astrophysical Journal Letters, 960, L3, \dodoi{10.3847/2041-8213/ad1691}

\bibitem[{Cadieux {et~al.}(2024{\natexlab{b}})Cadieux, Doyon, MacDonald,
  Turbet, Étienne Artigau, Lim, Radica, Fauchez, Salhi, Dang, Albert,
  Coulombe, Cowan, Lafrenière, L’Heureux, Piaulet-Ghorayeb, Benneke,
  Cloutier, Charnay, Cook, Fournier-Tondreau, Plotnykov, \&
  Valencia}]{Cadieux_JWST_2024}
Cadieux, C., Doyon, R., MacDonald, R.~J., {et~al.} 2024{\natexlab{b}}, The
  Astrophysical Journal Letters, 970, L2, \dodoi{10.3847/2041-8213/ad5afa}

\bibitem[{{Charbonneau} {et~al.}(2000){Charbonneau}, {Brown}, {Latham}, \&
  {Mayor}}]{Charbonneau2000}
{Charbonneau}, D., {Brown}, T.~M., {Latham}, D.~W., \& {Mayor}, M. 2000, \apjl,
  529, L45, \dodoi{10.1086/312457}

\bibitem[{{Charbonneau} {et~al.}(2009){Charbonneau}, {Berta}, {Irwin}, {Burke},
  {Nutzman}, {Buchhave}, {Lovis}, {Bonfils}, {Latham}, {Udry}, {Murray-Clay},
  {Holman}, {Falco}, {Winn}, {Queloz}, {Pepe}, {Mayor}, {Delfosse}, \&
  {Forveille}}]{Charbonneau2009}
{Charbonneau}, D., {Berta}, Z.~K., {Irwin}, J., {et~al.} 2009, \nat, 462, 891,
  \dodoi{10.1038/nature08679}

\bibitem[{{Chassefi{\`e}re} {et~al.}(2007){Chassefi{\`e}re}, {Leblanc}, \&
  {Langlais}}]{Chassefiere2007}
{Chassefi{\`e}re}, E., {Leblanc}, F., \& {Langlais}, B. 2007, \planss, 55, 343,
  \dodoi{10.1016/j.pss.2006.02.003}

\bibitem[{{Chaturvedi, P.} {et~al.}(2022){Chaturvedi, P.}, {Bluhm, P.}, {Nagel,
  E.}, {Hatzes, A. P.}, {Morello, G.}, {Brady, M.}, {Korth, J.},
  {Molaverdikhani, K.}, {Kossakowski, D.}, {Caballero, J. A.}, {Guenther, E.
  W.}, {Pall\'e, E.}, {Espinoza, N.}, {Seifahrt, A.}, {Lodieu, N.}, {Cifuentes,
  C.}, {Furlan, E.}, {Amado, P. J.}, {Barclay, T.}, {Bean, J.}, {B\'ejar, V. J.
  S.}, {Bergond, G.}, {Boyle, A. W.}, {Ciardi, D.}, {Collins, K. A.}, {Collins,
  K. I.}, {Esparza-Borges, E.}, {Fukui, A.}, {Gnilka, C. L.}, {Goeke, R.},
  {Guerra, P.}, {Henning, Th.}, {Herrero, E.}, {Howell, S. B.}, {Jeffers, S.
  V.}, {Jenkins, J. M.}, {Jensen, E. L. N.}, {Kasper, D.}, {Kodama, T.},
  {Latham, D. W.}, {L\'opez-Gonz\'alez, M. J.}, {Luque, R.}, {Montes, D.},
  {Morales, J. C.}, {Mori, M.}, {Murgas, F.}, {Narita, N.}, {Nowak, G.},
  {Parviainen, H.}, {Passegger, V. M.}, {Quirrenbach, A.}, {Reffert, S.},
  {Reiners, A.}, {Ribas, I.}, {Ricker, G. R.}, {Rodriguez, E.},
  {Rodr\'{\i}guez-L\'opez, C.}, {Schlecker, M.}, {Schwarz, R. P.}, {Schweitzer,
  A.}, {Seager, S.}, {Stef\'ansson, G.}, {Stockdale, C.}, {Tal-Or, L.},
  {Twicken, J. D.}, {Vanaverbeke, S.}, {Wang, G.}, {Watanabe, D.}, {Winn, J.
  N.}, \& {Zechmeister, M.}}]{chaturvedi_2022}
{Chaturvedi, P.}, {Bluhm, P.}, {Nagel, E.}, {et~al.} 2022, A\&A, 666, A155,
  \dodoi{10.1051/0004-6361/202244056}

\bibitem[{Cohen {et~al.}(2020)Cohen, Garraffo, Moschou, Drake,
  Alvarado-G{\'{o}}mez, Glocer, \& Fraschetti}]{Cohen_2020}
Cohen, O., Garraffo, C., Moschou, S.-P., {et~al.} 2020, The Astrophysical
  Journal, 897, 101, \dodoi{10.3847/1538-4357/ab9637}

\bibitem[{Damiano {et~al.}(2024)Damiano, Bello-Arufe, Yang, \&
  Hu}]{Damiano_2024}
Damiano, M., Bello-Arufe, A., Yang, J., \& Hu, R. 2024, The Astrophysical
  Journal Letters, 968, L22, \dodoi{10.3847/2041-8213/ad5204}

\bibitem[{Dittmann {et~al.}(2017)Dittmann, Irwin, Charbonneau, Bonfils,
  Astudillo-Defru, Haywood, Berta-Thompson, Newton, Rodriguez, Winters, Tan,
  Almenara, Bouchy, Delfosse, Forveille, Lovis, Murgas, Pepe, Santos, \&
  Dressing}]{Dittmann_2017}
Dittmann, J., Irwin, J., Charbonneau, D., {et~al.} 2017, Nature, 544, 333,
  \dodoi{10.1038/nature22055}

\bibitem[{{Dorn} {et~al.}(2015){Dorn}, {Khan}, {Heng}, {Connolly}, {Alibert},
  {Benz}, \& {Tackley}}]{Dorn2015}
{Dorn}, C., {Khan}, A., {Heng}, K., {et~al.} 2015, \aap, 577, A83,
  \dodoi{10.1051/0004-6361/201424915}

\bibitem[{{Dorn, Caroline} {et~al.}(2017){Dorn, Caroline}, {Venturini, Julia},
  {Khan, Amir}, {Heng, Kevin}, {Alibert, Yann}, {Helled, Ravit}, {Rivoldini,
  Attilio}, \& {Benz, Willy}}]{Dorn2017}
{Dorn, Caroline}, {Venturini, Julia}, {Khan, Amir}, {et~al.} 2017, A\&A, 597,
  A37, \dodoi{10.1051/0004-6361/201628708}

\bibitem[{{Dressing} {et~al.}(2019){Dressing}, {Hardegree-Ullman}, {Schlieder},
  {Newton}, {Vanderburg}, {Feinstein}, {Duvvuri}, {Arnold}, {Bristow},
  {Thackeray}, {Schwab Abrahams}, {Ciardi}, {Crossfield}, {Yu}, {Martinez},
  {Christiansen}, {Crepp}, \& {Isaacson}}]{Dressing_2019}
{Dressing}, C.~D., {Hardegree-Ullman}, K., {Schlieder}, J.~E., {et~al.} 2019,
  \aj, 158, 87, \dodoi{10.3847/1538-3881/ab2895}

\bibitem[{{Driscoll} \& {Barnes}(2015)}]{Driscoll2015}
{Driscoll}, P.~E., \& {Barnes}, R. 2015, Astrobiology, 15, 739,
  \dodoi{10.1089/ast.2015.1325}

\bibitem[{{Edwards} {et~al.}(2021){Edwards}, {Changeat}, {Mori}, {Anisman},
  {Morvan}, {Yip}, {Tsiaras}, {Al-Refaie}, {Waldmann}, \&
  {Tinetti}}]{Edwards_2021}
{Edwards}, B., {Changeat}, Q., {Mori}, M., {et~al.} 2021, \aj, 161, 44,
  \dodoi{10.3847/1538-3881/abc6a5}

\bibitem[{{Elkins-Tanton} \& {Seager}(2008)}]{2008ApJ...688..628E}
{Elkins-Tanton}, L.~T., \& {Seager}, S. 2008, \apj, 688, 628,
  \dodoi{10.1086/592316}

\bibitem[{{Fabrycky}(2009)}]{Fabrycky2009}
{Fabrycky}, D.~C. 2009, in IAU Symposium, Vol. 253, Transiting Planets, ed.
  F.~{Pont}, D.~{Sasselov}, \& M.~J. {Holman}, 173--179,
  \dodoi{10.1017/S1743921308026380}

\bibitem[{{Fauchez} {et~al.}(2019){Fauchez}, {Turbet}, {Villanueva}, {Wolf},
  {Arney}, {Kopparapu}, {Lincowski}, {Mandell}, {de Wit}, {Pidhorodetska},
  {Domagal-Goldman}, \& {Stevenson}}]{Fauchez_2019}
{Fauchez}, T.~J., {Turbet}, M., {Villanueva}, G.~L., {et~al.} 2019, \apj, 887,
  194, \dodoi{10.3847/1538-4357/ab5862}

\bibitem[{Fulton {et~al.}(2017)Fulton, Petigura, Howard, Isaacson, Marcy,
  Cargile, Hebb, Weiss, Johnson, Morton, Sinukoff, Crossfield, \&
  Hirsch}]{Fulton_2017}
Fulton, B.~J., Petigura, E.~A., Howard, A.~W., {et~al.} 2017, The Astronomical
  Journal, 154, 109, \dodoi{10.3847/1538-3881/aa80eb}

\bibitem[{{Gaia Collaboration} {et~al.}(2018){Gaia Collaboration}, {Brown, A.
  G. A.}, {Vallenari, A.}, {Prusti, T.}, {de Bruijne, J. H. J.}, {Babusiaux,
  C.}, {Bailer-Jones, C. A. L.}, {Biermann, M.}, {Evans, D. W.}, {Eyer, L.},
  {Jansen, F.}, {Jordi, C.}, {Klioner, S. A.}, {Lammers, U.}, {Lindegren, L.},
  {Luri, X.}, {Mignard, F.}, {Panem, C.}, {Pourbaix, D.}, {Randich, S.},
  {Sartoretti, P.}, {Siddiqui, H. I.}, {Soubiran, C.}, {van Leeuwen, F.},
  {Walton, N. A.}, {Arenou, F.}, {Bastian, U.}, {Cropper, M.}, {Drimmel, R.},
  {Katz, D.}, {Lattanzi, M. G.}, {Bakker, J.}, {Cacciari, C.}, {Casta\~neda,
  J.}, {Chaoul, L.}, {Cheek, N.}, {De Angeli, F.}, {Fabricius, C.}, {Guerra,
  R.}, {Holl, B.}, {Masana, E.}, {Messineo, R.}, {Mowlavi, N.}, {Nienartowicz,
  K.}, {Panuzzo, P.}, {Portell, J.}, {Riello, M.}, {Seabroke, G. M.}, {Tanga,
  P.}, {Th\'evenin, F.}, {Gracia-Abril, G.}, {Comoretto, G.},
  {Garcia-Reinaldos, M.}, {Teyssier, D.}, {Altmann, M.}, {Andrae, R.}, {Audard,
  M.}, {Bellas-Velidis, I.}, {Benson, K.}, {Berthier, J.}, {Blomme, R.},
  {Burgess, P.}, {Busso, G.}, {Carry, B.}, {Cellino, A.}, {Clementini, G.},
  {Clotet, M.}, {Creevey, O.}, {Davidson, M.}, {De Ridder, J.}, {Delchambre,
  L.}, {Dell\'{}Oro, A.}, {Ducourant, C.}, {Fern\'andez-Hern\'andez, J.},
  {Fouesneau, M.}, {Fr\'emat, Y.}, {Galluccio, L.}, {Garc\'{\i}a-Torres, M.},
  {Gonz\'alez-N\'u\~nez, J.}, {Gonz\'alez-Vidal, J. J.}, {Gosset, E.}, {Guy, L.
  P.}, {Halbwachs, J.-L.}, {Hambly, N. C.}, {Harrison, D. L.}, {Hern\'andez,
  J.}, {Hestroffer, D.}, {Hodgkin, S. T.}, {Hutton, A.}, {Jasniewicz, G.},
  {Jean-Antoine-Piccolo, A.}, {Jordan, S.}, {Korn, A. J.}, {Krone-Martins, A.},
  {Lanzafame, A. C.}, {Lebzelter, T.}, {L\"offler, W.}, {Manteiga, M.},
  {Marrese, P. M.}, {Mart\'{\i}n-Fleitas, J. M.}, {Moitinho, A.}, {Mora, A.},
  {Muinonen, K.}, {Osinde, J.}, {Pancino, E.}, {Pauwels, T.}, {Petit, J.-M.},
  {Recio-Blanco, A.}, {Richards, P. J.}, {Rimoldini, L.}, {Robin, A. C.},
  {Sarro, L. M.}, {Siopis, C.}, {Smith, M.}, {Sozzetti, A.}, {S\"uveges, M.},
  {Torra, J.}, {van Reeven, W.}, {Abbas, U.}, {Abreu Aramburu, A.}, {Accart,
  S.}, {Aerts, C.}, {Altavilla, G.}, {\'Alvarez, M. A.}, {Alvarez, R.}, {Alves,
  J.}, {Anderson, R. I.}, {Andrei, A. H.}, {Anglada Varela, E.}, {Antiche, E.},
  {Antoja, T.}, {Arcay, B.}, {Astraatmadja, T. L.}, {Bach, N.}, {Baker, S. G.},
  {Balaguer-N\'u\~nez, L.}, {Balm, P.}, {Barache, C.}, {Barata, C.}, {Barbato,
  D.}, {Barblan, F.}, {Barklem, P. S.}, {Barrado, D.}, {Barros, M.}, {Barstow,
  M. A.}, {Bartholom\'e Mu\~noz, S.}, {Bassilana, J.-L.}, {Becciani, U.},
  {Bellazzini, M.}, {Berihuete, A.}, {Bertone, S.}, {Bianchi, L.}, {Bienaym\'e,
  O.}, {Blanco-Cuaresma, S.}, {Boch, T.}, {Boeche, C.}, {Bombrun, A.},
  {Borrachero, R.}, {Bossini, D.}, {Bouquillon, S.}, {Bourda, G.}, {Bragaglia,
  A.}, {Bramante, L.}, {Breddels, M. A.}, {Bressan, A.}, {Brouillet, N.},
  {Br\"usemeister, T.}, {Brugaletta, E.}, {Bucciarelli, B.}, {Burlacu, A.},
  {Busonero, D.}, {Butkevich, A. G.}, {Buzzi, R.}, {Caffau, E.}, {Cancelliere,
  R.}, {Cannizzaro, G.}, {Cantat-Gaudin, T.}, {Carballo, R.}, {Carlucci, T.},
  {Carrasco, J. M.}, {Casamiquela, L.}, {Castellani, M.}, {Castro-Ginard, A.},
  {Charlot, P.}, {Chemin, L.}, {Chiavassa, A.}, {Cocozza, G.}, {Costigan, G.},
  {Cowell, S.}, {Crifo, F.}, {Crosta, M.}, {Crowley, C.}, {Cuypers+, J.},
  {Dafonte, C.}, {Damerdji, Y.}, {Dapergolas, A.}, {David, P.}, {David, M.},
  {de Laverny, P.}, {De Luise, F.}, {De March, R.}, {de Martino, D.}, {de
  Souza, R.}, {de Torres, A.}, {Debosscher, J.}, {del Pozo, E.}, {Delbo, M.},
  {Delgado, A.}, {Delgado, H. E.}, {Di Matteo, P.}, {Diakite, S.}, {Diener,
  C.}, {Distefano, E.}, {Dolding, C.}, {Drazinos, P.}, {Dur\'an, J.},
  {Edvardsson, B.}, {Enke, H.}, {Eriksson, K.}, {Esquej, P.}, {Eynard Bontemps,
  G.}, {Fabre, C.}, {Fabrizio, M.}, {Faigler, S.}, {Falc\~ao, A. J.}, {Farr\`as
  Casas, M.}, {Federici, L.}, {Fedorets, G.}, {Fernique, P.}, {Figueras, F.},
  {Filippi, F.}, {Findeisen, K.}, {Fonti, A.}, {Fraile, E.}, {Fraser, M.},
  {Fr\'ezouls, B.}, {Gai, M.}, {Galleti, S.}, {Garabato, D.},
  {Garc\'{\i}a-Sedano, F.}, {Garofalo, A.}, {Garralda, N.}, {Gavel, A.},
  {Gavras, P.}, {Gerssen, J.}, {Geyer, R.}, {Giacobbe, P.}, {Gilmore, G.},
  {Girona, S.}, {Giuffrida, G.}, {Glass, F.}, {Gomes, M.}, {Granvik, M.},
  {Gueguen, A.}, {Guerrier, A.}, {Guiraud, J.}, {Guti\'errez-S\'anchez, R.},
  {Haigron, R.}, {Hatzidimitriou, D.}, {Hauser, M.}, {Haywood, M.}, {Heiter,
  U.}, {Helmi, A.}, {Heu, J.}, {Hilger, T.}, {Hobbs, D.}, {Hofmann, W.},
  {Holland, G.}, {Huckle, H. E.}, {Hypki, A.}, {Icardi, V.}, {Jan\ss{}en, K.},
  {Jevardat de Fombelle, G.}, {Jonker, P. G.}, {Juh\'asz, \'A. L.}, {Julbe,
  F.}, {Karampelas, A.}, {Kewley, A.}, {Klar, J.}, {Kochoska, A.}, {Kohley,
  R.}, {Kolenberg, K.}, {Kontizas, M.}, {Kontizas, E.}, {Koposov, S. E.},
  {Kordopatis, G.}, {Kostrzewa-Rutkowska, Z.}, {Koubsky, P.}, {Lambert, S.},
  {Lanza, A. F.}, {Lasne, Y.}, {Lavigne, J.-B.}, {Le Fustec, Y.}, {Le
  Poncin-Lafitte, C.}, {Lebreton, Y.}, {Leccia, S.}, {Leclerc, N.},
  {Lecoeur-Taibi, I.}, {Lenhardt, H.}, {Leroux, F.}, {Liao, S.}, {Licata, E.},
  {Lindstr\o{}m, H. E. P.}, {Lister, T. A.}, {Livanou, E.}, {Lobel, A.},
  {L\'opez, M.}, {Managau, S.}, {Mann, R. G.}, {Mantelet, G.}, {Marchal, O.},
  {Marchant, J. M.}, {Marconi, M.}, {Marinoni, S.}, {Marschalk\'o, G.},
  {Marshall, D. J.}, {Martino, M.}, {Marton, G.}, {Mary, N.}, {Massari, D.},
  {Matijevic, G.}, {Mazeh, T.}, {McMillan, P. J.}, {Messina, S.}, {Michalik,
  D.}, {Millar, N. R.}, {Molina, D.}, {Molinaro, R.}, {Moln\'ar, L.},
  {Montegriffo, P.}, {Mor, R.}, {Morbidelli, R.}, {Morel, T.}, {Morris, D.},
  {Mulone, A. F.}, {Muraveva, T.}, {Musella, I.}, {Nelemans, G.}, {Nicastro,
  L.}, {Noval, L.}, {O\'{}Mullane, W.}, {Ord\'enovic, C.}, {Ord\'o\~nez-Blanco,
  D.}, {Osborne, P.}, {Pagani, C.}, {Pagano, I.}, {Pailler, F.}, {Palacin, H.},
  {Palaversa, L.}, {Panahi, A.}, {Pawlak, M.}, {Piersimoni, A. M.}, {Pineau,
  F.-X.}, {Plachy, E.}, {Plum, G.}, {Poggio, E.}, {Poujoulet, E.}, {Prsa, A.},
  {Pulone, L.}, {Racero, E.}, {Ragaini, S.}, {Rambaux, N.}, {Ramos-Lerate, M.},
  {Regibo, S.}, {Reyl\'e, C.}, {Riclet, F.}, {Ripepi, V.}, {Riva, A.}, {Rivard,
  A.}, {Rixon, G.}, {Roegiers, T.}, {Roelens, M.}, {Romero-G\'omez, M.},
  {Rowell, N.}, {Royer, F.}, {Ruiz-Dern, L.}, {Sadowski, G.}, {Sagrist\`a
  Sell\'es, T.}, {Sahlmann, J.}, {Salgado, J.}, {Salguero, E.}, {Sanna, N.},
  {Santana-Ros, T.}, {Sarasso, M.}, {Savietto, H.}, {Schultheis, M.}, {Sciacca,
  E.}, {Segol, M.}, {Segovia, J. C.}, {S\'egransan, D.}, {Shih, I-C.},
  {Siltala, L.}, {Silva, A. F.}, {Smart, R. L.}, {Smith, K. W.}, {Solano, E.},
  {Solitro, F.}, {Sordo, R.}, {Soria Nieto, S.}, {Souchay, J.}, {Spagna, A.},
  {Spoto, F.}, {Stampa, U.}, {Steele, I. A.}, {Steidelm\"uller, H.},
  {Stephenson, C. A.}, {Stoev, H.}, {Suess, F. F.}, {Surdej, J.}, {Szabados,
  L.}, {Szegedi-Elek, E.}, {Tapiador, D.}, {Taris, F.}, {Tauran, G.}, {Taylor,
  M. B.}, {Teixeira, R.}, {Terrett, D.}, {Teyssandier, P.}, {Thuillot, W.},
  {Titarenko, A.}, {Torra Clotet, F.}, {Turon, C.}, {Ulla, A.}, {Utrilla, E.},
  {Uzzi, S.}, {Vaillant, M.}, {Valentini, G.}, {Valette, V.}, {van Elteren,
  A.}, {Van Hemelryck, E.}, {van Leeuwen, M.}, {Vaschetto, M.}, {Vecchiato,
  A.}, {Veljanoski, J.}, {Viala, Y.}, {Vicente, D.}, {Vogt, S.}, {von Essen,
  C.}, {Voss, H.}, {Votruba, V.}, {Voutsinas, S.}, {Walmsley, G.}, {Weiler,
  M.}, {Wertz, O.}, {Wevers, T.}, {Wyrzykowski, L.}, {Yoldas, A.}, {Zerjal,
  M.}, {Ziaeepour, H.}, {Zorec, J.}, {Zschocke, S.}, {Zucker, S.}, {Zurbach,
  C.}, \& {Zwitter, T.}}]{Gaia_Collaboration_2018}
{Gaia Collaboration}, {Brown, A. G. A.}, {Vallenari, A.}, {et~al.} 2018, A\&A,
  616, A1, \dodoi{10.1051/0004-6361/201833051}

\bibitem[{{Gaia Collaboration} {et~al.}(2021){Gaia Collaboration}, {Brown},
  {Vallenari}, {Prusti}, {de Bruijne}, {Babusiaux}, {Biermann}, {Creevey},
  {Evans}, {Eyer}, \& et~al.}]{Gaia_Collaboration_2020}
{Gaia Collaboration}, {Brown}, A.~G.~A., {Vallenari}, A., {et~al.} 2021, \aap,
  649, A1, \dodoi{10.1051/0004-6361/202039657}

\bibitem[{{Gardner} {et~al.}(2006){Gardner}, {Mather}, {Clampin}, {Doyon},
  {Greenhouse}, {Hammel}, {Hutchings}, {Jakobsen}, {Lilly}, {Long}, {Lunine},
  {McCaughrean}, {Mountain}, {Nella}, {Rieke}, {Rieke}, {Rix}, {Smith},
  {Sonneborn}, {Stiavelli}, {Stockman}, {Windhorst}, \&
  {Wright}}]{GardnerJWST2006}
{Gardner}, J.~P., {Mather}, J.~C., {Clampin}, M., {et~al.} 2006, \ssr, 123,
  485, \dodoi{10.1007/s11214-006-8315-7}

\bibitem[{{Gilli} {et~al.}(2006){Gilli}, {Israelian}, {Ecuvillon}, {Santos}, \&
  {Mayor}}]{Gilli2006}
{Gilli}, G., {Israelian}, G., {Ecuvillon}, A., {Santos}, N.~C., \& {Mayor}, M.
  2006, \aap, 449, 723, \dodoi{10.1051/0004-6361:20053850}

\bibitem[{Gillmann {et~al.}(2024)Gillmann, Hakim, Lourenço, Quanz, \&
  Sossi}]{Gillmann2024}
Gillmann, C., Hakim, K., Lourenço, D., Quanz, S.~P., \& Sossi, P.~A. 2024,
  Space: Science \&amp; Technology, 4, 0075, \dodoi{10.34133/space.0075}

\bibitem[{{Goldblatt} \& {Watson}(2012)}]{Goldblatt2012}
{Goldblatt}, C., \& {Watson}, A.~J. 2012, Philosophical Transactions of the
  Royal Society of London Series A, 370, 4197, \dodoi{10.1098/rsta.2012.0004}

\bibitem[{{Gomes} \& {Ferraz-Mello}(2020)}]{Gomes2020}
{Gomes}, G.~O., \& {Ferraz-Mello}, S. 2020, \mnras, 494, 5082,
  \dodoi{10.1093/mnras/staa1110}

\bibitem[{Grande {et~al.}(2022)Grande, Pham, Smith, Boisvert, Huang, Smith,
  Goldman, Belof, Tschauner, Steffen, \& Salamat}]{Grande2022}
Grande, Z.~M., Pham, C.~H., Smith, D., {et~al.} 2022, Phys. Rev. B, 105,
  104109, \dodoi{10.1103/PhysRevB.105.104109}

\bibitem[{{Grasset} {et~al.}(2009){Grasset}, {Schneider}, \&
  {Sotin}}]{Grasset2009}
{Grasset}, O., {Schneider}, J., \& {Sotin}, C. 2009, \apj, 693, 722,
  \dodoi{10.1088/0004-637X/693/1/722}

\bibitem[{Guinan \& Engle(2019)}]{Guinan_2019}
Guinan, E.~F., \& Engle, S.~G. 2019, Research Notes of the AAS, 3, 189,
  \dodoi{10.3847/2515-5172/ab6086}

\bibitem[{{Haldemann} {et~al.}(2024){Haldemann}, {Dorn}, {Venturini},
  {Alibert}, \& {Benz}}]{HaldemannBiceps2024}
{Haldemann}, J., {Dorn}, C., {Venturini}, J., {Alibert}, Y., \& {Benz}, W.
  2024, \aap, 681, A96, \dodoi{10.1051/0004-6361/202346965}

\bibitem[{Hammond \& Pierrehumbert(2017)}]{Hammond_2017}
Hammond, M., \& Pierrehumbert, R.~T. 2017, The Astrophysical Journal, 849, 152,
  \dodoi{10.3847/1538-4357/aa9328}

\bibitem[{Haqq-Misra {et~al.}(2018)Haqq-Misra, Wolf, Joshi, Zhang, \&
  Kopparapu}]{Haqq-Misra_2018}
Haqq-Misra, J., Wolf, E.~T., Joshi, M., Zhang, X., \& Kopparapu, R.~K. 2018,
  The Astrophysical Journal, 852, 67, \dodoi{10.3847/1538-4357/aa9f1f}

\bibitem[{{Hart}(1978)}]{Hart1978}
{Hart}, M.~H. 1978, \icarus, 33, 23, \dodoi{10.1016/0019-1035(78)90021-0}

\bibitem[{Henning {et~al.}(2009)Henning, O{\textquotesingle}Connell, \&
  Sasselov}]{Henning_2009}
Henning, W.~G., O{\textquotesingle}Connell, R.~J., \& Sasselov, D.~D. 2009, The
  Astrophysical Journal, 707, 1000, \dodoi{10.1088/0004-637x/707/2/1000}

\bibitem[{Hinkel \& Unterborn(2018)}]{Hinkel_2018}
Hinkel, N.~R., \& Unterborn, C.~T. 2018, The Astrophysical Journal, 853, 83,
  \dodoi{10.3847/1538-4357/aaa5b4}

\bibitem[{Huang {et~al.}(2022)Huang, Rice, \& Steffen}]{MagratheaHuang2022}
Huang, C., Rice, D.~R., \& Steffen, J.~H. 2022, Monthly Notices of the Royal
  Astronomical Society, 513, 5256, \dodoi{10.1093/mnras/stac1133}

\bibitem[{{Huang}(1959)}]{Huang1959A}
{Huang}, S.-S. 1959, American Scientist, 47, 397

\bibitem[{{Johansen} {et~al.}(2023){Johansen}, {Ronnet}, {Schiller}, {Deng}, \&
  {Bizzarro}}]{2023A&A...671A..74J}
{Johansen}, A., {Ronnet}, T., {Schiller}, M., {Deng}, Z., \& {Bizzarro}, M.
  2023, \aap, 671, A74, \dodoi{10.1051/0004-6361/202142141}

\bibitem[{Kaltenegger \& Sasselov(2011)}]{Kaltenegger_2011}
Kaltenegger, L., \& Sasselov, D. 2011, The Astrophysical Journal Letters, 736,
  L25, \dodoi{10.1088/2041-8205/736/2/L25}

\bibitem[{Karato(2015)}]{Karato2015}
Karato, S. 2015, in Treatise on Geophysics (Second Edition), second edition
  edn., ed. G.~Schubert (Oxford: Elsevier), 105--144,
  \dodoi{https://doi.org/10.1016/B978-0-444-53802-4.00156-1}

\bibitem[{{Kasting} \& {Catling}(2003)}]{Kasting2003}
{Kasting}, J.~F., \& {Catling}, D. 2003, \araa, 41, 429,
  \dodoi{10.1146/annurev.astro.41.071601.170049}

\bibitem[{{Kasting} {et~al.}(1993){Kasting}, {Whitmire}, \&
  {Reynolds}}]{Kasting1993}
{Kasting}, J.~F., {Whitmire}, D.~P., \& {Reynolds}, R.~T. 1993, \icarus, 101,
  108, \dodoi{10.1006/icar.1993.1010}

\bibitem[{{Katayama} \& {Karato}(2008)}]{Katayama2008}
{Katayama}, I., \& {Karato}, S.-I. 2008, Physics of the Earth and Planetary
  Interiors, 166, 57, \dodoi{10.1016/j.pepi.2007.10.004}

\bibitem[{{Konatham} {et~al.}(2020){Konatham}, {Martin-Torres}, \&
  {Zorzano}}]{Konatham_2020}
{Konatham}, S., {Martin-Torres}, J., \& {Zorzano}, M.-P. 2020, Proceedings of
  the Royal Society of London Series A, 476, 20200148,
  \dodoi{10.1098/rspa.2020.0148}

\bibitem[{Kopparapu {et~al.}(2013)Kopparapu, Ramirez, Kasting, Eymet, Robinson,
  Mahadevan, Terrien, Domagal-Goldman, Meadows, \& Deshpande}]{Kopparapu_2013}
Kopparapu, R.~K., Ramirez, R., Kasting, J.~F., {et~al.} 2013, The Astrophysical
  Journal, 765, 131, \dodoi{10.1088/0004-637X/765/2/131}

\bibitem[{{Kramm, U.} {et~al.}(2012){Kramm, U.}, {Nettelmann, N.}, {Fortney, J.
  J.}, {Neuh\"auser, R.}, \& {Redmer, R.}}]{Kramm_2011}
{Kramm, U.}, {Nettelmann, N.}, {Fortney, J. J.}, {Neuh\"auser, R.}, \& {Redmer,
  R.} 2012, A\&A, 538, A146, \dodoi{10.1051/0004-6361/201118141}

\bibitem[{kumar Kopparapu {et~al.}(2017)kumar Kopparapu, Wolf, Arney, Batalha,
  Haqq-Misra, Grimm, \& Heng}]{Kopparapu_2017}
kumar Kopparapu, R., Wolf, E.~T., Arney, G., {et~al.} 2017, The Astrophysical
  Journal, 845, 5, \dodoi{10.3847/1538-4357/aa7cf9}

\bibitem[{kumar Kopparapu {et~al.}(2016)kumar Kopparapu, Wolf, Haqq-Misra,
  Yang, Kasting, Meadows, Terrien, \& Mahadevan}]{Kopparapu_2016}
kumar Kopparapu, R., Wolf, E.~T., Haqq-Misra, J., {et~al.} 2016, The
  Astrophysical Journal, 819, 84, \dodoi{10.3847/0004-637X/819/1/84}

\bibitem[{{Lazio} {et~al.}(2019){Lazio}, {Hallinan}, {Airapetian}, {Brain},
  {Clarke}, {Dolch}, {Dong}, {Driscoll}, {Fares}, {Griessmeier}, {Farrell},
  {Kasper}, {Murphy}, {Rogers}, {Shkolnik}, {Stanley}, {Strugarek}, {Turner},
  {Wolszczan}, {Zarka}, {Knapp}, {Lynch}, \& {Turner}}]{Lazio2019}
{Lazio}, J., {Hallinan}, G., {Airapetian}, A., {et~al.} 2019, \baas, 51, 135,
  \dodoi{10.48550/arXiv.1803.06487}

\bibitem[{Leconte {et~al.}(2015)Leconte, Wu, Menou, \& Murray}]{leconte2015}
Leconte, J., Wu, H., Menou, K., \& Murray, N. 2015, Science, 347, 632,
  \dodoi{10.1126/science.1258686}

\bibitem[{{Lillo-Box, J.} {et~al.}(2020){Lillo-Box, J.}, {Figueira, P.},
  {Leleu, A.}, {Acu\~na, L.}, {Faria, J. P.}, {Hara, N.}, {Santos, N. C.},
  {Correia, A. C. M.}, {Robutel, P.}, {Deleuil, M.}, {Barrado, D.}, {Sousa,
  S.}, {Bonfils, X.}, {Mousis, O.}, {Almenara, J. M.}, {Astudillo-Defru, N.},
  {Marcq, E.}, {Udry, S.}, {Lovis, C.}, \& {Pepe, F.}}]{Lillo-Box_2020}
{Lillo-Box, J.}, {Figueira, P.}, {Leleu, A.}, {et~al.} 2020, A\&A, 642, A121,
  \dodoi{10.1051/0004-6361/202038922}

\bibitem[{{Lodders}(2003)}]{lodders2003}
{Lodders}, K. 2003, \apj, 591, 1220, \dodoi{10.1086/375492}

\bibitem[{{Lundin} {et~al.}(2007){Lundin}, {Lammer}, \& {Ribas}}]{Lundin2007}
{Lundin}, R., {Lammer}, H., \& {Ribas}, I. 2007, \ssr, 129, 245,
  \dodoi{10.1007/s11214-007-9176-4}

\bibitem[{Luque \& Pallé(2022)}]{Luque_2022}
Luque, R., \& Pallé, E. 2022, Science, 377, 1211,
  \dodoi{10.1126/science.abl7164}

\bibitem[{{Lynch} {et~al.}(2017){Lynch}, {Murphy}, {Kaplan}, {Ireland}, \&
  {Bell}}]{Lynch2017}
{Lynch}, C.~R., {Murphy}, T., {Kaplan}, D.~L., {Ireland}, M., \& {Bell}, M.~E.
  2017, \mnras, 467, 3447, \dodoi{10.1093/mnras/stx354}

\bibitem[{Madhusudhan {et~al.}(2021)Madhusudhan, Piette, \&
  Constantinou}]{Madhusudhan_2021}
Madhusudhan, N., Piette, A. A.~A., \& Constantinou, S. 2021, The Astrophysical
  Journal, 918, 1, \dodoi{10.3847/1538-4357/abfd9c}

\bibitem[{Madhusudhan {et~al.}(2023)Madhusudhan, Sarkar, Constantinou,
  Holmberg, Piette, \& Moses}]{Madhusudhan_2023}
Madhusudhan, N., Sarkar, S., Constantinou, S., {et~al.} 2023, The Astrophysical
  Journal Letters, 956, L13, \dodoi{10.3847/2041-8213/acf577}

\bibitem[{{Marfil} {et~al.}(2021){Marfil}, {Tabernero}, {Montes}, {Caballero},
  {L{\'a}zaro}, {Gonz{\'a}lez Hern{\'a}ndez}, {Nagel}, {Passegger},
  {Schweitzer}, {Ribas}, {Reiners}, {Quirrenbach}, {Amado}, {Cifuentes},
  {Cort{\'e}s-Contreras}, {Dreizler}, {Duque-Arribas},
  {Galad{\'\i}-Enr{\'\i}quez}, {Henning}, {Jeffers}, {Kaminski}, {K{\"u}rster},
  {Lafarga}, {L{\'o}pez-Gallifa}, {Morales}, {Shan}, \&
  {Zechmeister}}]{Marfil_2021}
{Marfil}, E., {Tabernero}, H.~M., {Montes}, D., {et~al.} 2021, \aap, 656, A162,
  \dodoi{10.1051/0004-6361/202141980}

\bibitem[{{Mayor} {et~al.}(2003){Mayor}, {Pepe}, {Queloz}, {Bouchy},
  {Rupprecht}, {Lo Curto}, {Avila}, {Benz}, {Bertaux}, {Bonfils}, {Dall},
  {Dekker}, {Delabre}, {Eckert}, {Fleury}, {Gilliotte}, {Gojak}, {Guzman},
  {Kohler}, {Lizon}, {Longinotti}, {Lovis}, {Megevand}, {Pasquini}, {Reyes},
  {Sivan}, {Sosnowska}, {Soto}, {Udry}, {van Kesteren}, {Weber}, \&
  {Weilenmann}}]{2003Msngr.114...20M}
{Mayor}, M., {Pepe}, F., {Queloz}, D., {et~al.} 2003, The Messenger, 114, 20

\bibitem[{{Mazeh} {et~al.}(2000){Mazeh}, {Naef}, {Torres}, {Latham}, {Mayor},
  {Beuzit}, {Brown}, {Buchhave}, {Burnet}, {Carney}, {Charbonneau}, {Drukier},
  {Laird}, {Pepe}, {Perrier}, {Queloz}, {Santos}, {Sivan}, {Udry}, \&
  {Zucker}}]{2000ApJ...532L..55M}
{Mazeh}, T., {Naef}, D., {Torres}, G., {et~al.} 2000, \apjl, 532, L55,
  \dodoi{10.1086/312558}

\bibitem[{Montet {et~al.}(2015)Montet, Morton, Foreman-Mackey, Johnson, Hogg,
  Bowler, Latham, Bieryla, \& Mann}]{Montet_2015}
Montet, B.~T., Morton, T.~D., Foreman-Mackey, D., {et~al.} 2015, The
  Astrophysical Journal, 809, 25, \dodoi{10.1088/0004-637X/809/1/25}

\bibitem[{Nixon \& Madhusudhan(2021)}]{Nixon2021}
Nixon, M.~C., \& Madhusudhan, N. 2021, Monthly Notices of the Royal
  Astronomical Society, 505, 3414, \dodoi{10.1093/mnras/stab1500}

\bibitem[{{Oganov}(2004)}]{Oganov2004}
{Oganov}, A.~R. 2004, in APS March Meeting Abstracts, APS Meeting Abstracts,
  A11.001

\bibitem[{{Peale} \& {Cassen}(1978)}]{Peale_1978}
{Peale}, S.~J., \& {Cassen}, P. 1978, \icarus, 36, 245,
  \dodoi{10.1016/0019-1035(78)90109-4}

\bibitem[{Peale {et~al.}(1979)Peale, Cassen, \& Reynolds}]{Peale_1979}
Peale, S.~J., Cassen, P., \& Reynolds, R.~T. 1979, Science, 203, 892,
  \dodoi{10.1126/science.203.4383.892}

\bibitem[{{Perdelwitz} {et~al.}(2024){Perdelwitz}, {Trifonov}, {Teklu},
  {Sreenivas}, \& {Tal-Or}}]{Perdelwitz2024}
{Perdelwitz}, V., {Trifonov}, T., {Teklu}, J.~T., {Sreenivas}, K.~R., \&
  {Tal-Or}, L. 2024, \aap, 683, A125, \dodoi{10.1051/0004-6361/202348263}

\bibitem[{{Perdelwitz} {et~al.}(2021){Perdelwitz}, {Mittag}, {Tal-Or},
  {Schmitt}, {Caballero}, {Jeffers}, {Reiners}, {Schweitzer}, {Trifonov},
  {Ribas}, {Quirrenbach}, {Amado}, {Seifert}, {Cifuentes},
  {Cort{\'e}s-Contreras}, {Montes}, {Revilla}, \&
  {Skrzypinski}}]{2021A&A...652A.116P}
{Perdelwitz}, V., {Mittag}, M., {Tal-Or}, L., {et~al.} 2021, \aap, 652, A116,
  \dodoi{10.1051/0004-6361/202140889}

\bibitem[{Rauscher {et~al.}(2008)Rauscher, Menou, Cho, Seager, \&
  Hansen}]{Rauscher_2008}
Rauscher, E., Menou, K., Cho, J. Y.-K., Seager, S., \& Hansen, B. M.~S. 2008,
  The Astrophysical Journal, 681, 1646, \dodoi{10.1086/589499}

\bibitem[{{Raymond} {et~al.}(2004){Raymond}, {Quinn}, \&
  {Lunine}}]{2004Icar..168....1R}
{Raymond}, S.~N., {Quinn}, T., \& {Lunine}, J.~I. 2004, \icarus, 168, 1,
  \dodoi{10.1016/j.icarus.2003.11.019}

\bibitem[{{Reid} {et~al.}(1995){Reid}, {Hawley}, \& {Gizis}}]{Reid_1995}
{Reid}, I.~N., {Hawley}, S.~L., \& {Gizis}, J.~E. 1995, \aj, 110, 1838,
  \dodoi{10.1086/117655}

\bibitem[{{Reiners, A.} {et~al.}(2018){Reiners, A.}, {Zechmeister, M.},
  {Caballero, J. A.}, {Ribas, I.}, {Morales, J. C.}, {Jeffers, S. V.},
  {Schöfer, P.}, {Tal-Or, L.}, {Quirrenbach, A.}, {Amado, P. J.}, {Kaminski,
  A.}, {Seifert, W.}, {Abril, M.}, {Aceituno, J.}, {Alonso-Floriano, F. J.},
  {Ammler-von Eiff, M.}, {Antona, R.}, {Anglada-Escudé, G.}, {Anwand-Heerwart,
  H.}, {Arroyo-Torres, B.}, {Azzaro, M.}, {Baroch, D.}, {Barrado, D.}, {Bauer,
  F. F.}, {Becerril, S.}, {Béjar, V. J. S.}, {Benítez, D.}, {Berdinas̃, Z.
  M.}, {Bergond, G.}, {Blümcke, M.}, {Brinkmöller, M.}, {del Burgo, C.},
  {Cano, J.}, {Cárdenas Vázquez, M. C.}, {Casal, E.}, {Cifuentes, C.},
  {Claret, A.}, {Colomé, J.}, {Cortés-Contreras, M.}, {Czesla, S.},
  {Díez-Alonso, E.}, {Dreizler, S.}, {Feiz, C.}, {Fernández, M.}, {Ferro, I.
  M.}, {Fuhrmeister, B.}, {Galadí-Enríquez, D.}, {Garcia-Piquer, A.},
  {García Vargas, M. L.}, {Gesa, L.}, {Gómez Galera, V.}, {González
  Hernández, J. I.}, {González-Peinado, R.}, {Grözinger, U.}, {Grohnert,
  S.}, {Guàrdia, J.}, {Guenther, E. W.}, {Guijarro, A.}, {de Guindos, E.},
  {Gutiérrez-Soto, J.}, {Hagen, H.-J.}, {Hatzes, A. P.}, {Hauschildt, P. H.},
  {Hedrosa, R. P.}, {Helmling, J.}, {Henning, Th.}, {Hermelo, I.}, {Hernández
  Arabí, R.}, {Hernández Castaño, L.}, {Hernández Hernando, F.}, {Herrero,
  E.}, {Huber, A.}, {Huke, P.}, {Johnson, E. N.}, {de Juan, E.}, {Kim, M.},
  {Klein, R.}, {Klüter, J.}, {Klutsch, A.}, {Kürster, M.}, {Lafarga, M.},
  {Lamert, A.}, {Lampón, M.}, {Lara, L. M.}, {Laun, W.}, {Lemke, U.}, {Lenzen,
  R.}, {Launhardt, R.}, {López del Fresno, M.}, {López-González, J.},
  {López-Puertas, M.}, {López Salas, J. F.}, {López-Santiago, J.}, {Luque,
  R.}, {Magán Madinabeitia, H.}, {Mall, U.}, {Mancini, L.}, {Mandel, H.},
  {Marfil, E.}, {Marín Molina, J. A.}, {Maroto Fernández, D.}, {Martín, E.
  L.}, {Martín-Ruiz, S.}, {Marvin, C. J.}, {Mathar, R. J.}, {Mirabet, E.},
  {Montes, D.}, {Moreno-Raya, M. E.}, {Moya, A.}, {Mundt, R.}, {Nagel, E.},
  {Naranjo, V.}, {Nortmann, L.}, {Nowak, G.}, {Ofir, A.}, {Oreiro, R.},
  {Pallé, E.}, {Panduro, J.}, {Pascual, J.}, {Passegger, V. M.}, {Pavlov, A.},
  {Pedraz, S.}, {Pérez-Calpena, A.}, {Pérez Medialdea, D.}, {Perger, M.},
  {Perryman, M. A. C.}, {Pluto, M.}, {Rabaza, O.}, {Ramón, A.}, {Rebolo, R.},
  {Redondo, P.}, {Reffert, S.}, {Reinhart, S.}, {Rhode, P.}, {Rix, H.-W.},
  {Rodler, F.}, {Rodríguez, E.}, {Rodríguez-López, C.}, {Rodríguez
  Trinidad, A.}, {Rohloff, R.-R.}, {Rosich, A.}, {Sadegi, S.},
  {Sánchez-Blanco, E.}, {Sánchez Carrasco, M. A.}, {Sánchez-López, A.},
  {Sanz-Forcada, J.}, {Sarkis, P.}, {Sarmiento, L. F.}, {Schäfer, S.},
  {Schmitt, J. H. M. M.}, {Schiller, J.}, {Schweitzer, A.}, {Solano, E.},
  {Stahl, O.}, {Strachan, J. B. P.}, {Stürmer, J.}, {Suárez, J. C.},
  {Tabernero, H. M.}, {Tala, M.}, {Trifonov, T.}, {Tulloch, S. M.}, {Ulbrich,
  R. G.}, {Veredas, G.}, {Vico Linares, J. I.}, {Vilardell, F.}, {Wagner, K.},
  {Winkler, J.}, {Wolthoff, V.}, {Xu, W.}, {Yan, F.}, \& {Zapatero Osorio, M.
  R.}}]{Reiners2018}
{Reiners, A.}, {Zechmeister, M.}, {Caballero, J. A.}, {et~al.} 2018, A\&A, 612,
  A49, \dodoi{10.1051/0004-6361/201732054}

\bibitem[{{Ricker} {et~al.}(2014){Ricker}, {Winn}, {Vanderspek}, {Latham},
  {Bakos}, {Bean}, {Berta-Thompson}, {Brown}, {Buchhave}, {Butler}, {Butler},
  {Chaplin}, {Charbonneau}, {Christensen-Dalsgaard}, {Clampin}, {Deming},
  {Doty}, {De Lee}, {Dressing}, {Dunham}, {Endl}, {Fressin}, {Ge}, {Henning},
  {Holman}, {Howard}, {Ida}, {Jenkins}, {Jernigan}, {Johnson}, {Kaltenegger},
  {Kawai}, {Kjeldsen}, {Laughlin}, {Levine}, {Lin}, {Lissauer}, {MacQueen},
  {Marcy}, {McCullough}, {Morton}, {Narita}, {Paegert}, {Palle}, {Pepe},
  {Pepper}, {Quirrenbach}, {Rinehart}, {Sasselov}, {Sato}, {Seager},
  {Sozzetti}, {Stassun}, {Sullivan}, {Szentgyorgyi}, {Torres}, {Udry}, \&
  {Villasenor}}]{RickerTESS2014}
{Ricker}, G.~R., {Winn}, J.~N., {Vanderspek}, R., {et~al.} 2014, in Society of
  Photo-Optical Instrumentation Engineers (SPIE) Conference Series, Vol. 9143,
  Space Telescopes and Instrumentation 2014: Optical, Infrared, and Millimeter
  Wave, ed. J.~{Oschmann}, Jacobus~M., M.~{Clampin}, G.~G. {Fazio}, \& H.~A.
  {MacEwen}, 914320, \dodoi{10.1117/12.2063489}

\bibitem[{{Rogers}(2015)}]{Rogers2015Radii}
{Rogers}, L.~A. 2015, \apj, 801, 41, \dodoi{10.1088/0004-637X/801/1/41}

\bibitem[{Rogers \& Seager(2010)}]{Rogers_2010}
Rogers, L. J.~A., \& Seager, S. 2010, The Astrophysical Journal, 712, 974

\bibitem[{Sakai {et~al.}(2016)Sakai, Dekura, \& Hirao}]{sakai2016experimental}
Sakai, T., Dekura, H., \& Hirao, N. 2016, Scientific Reports, 6, 22652,
  \dodoi{10.1038/srep22652}

\bibitem[{Sarkis {et~al.}(2018)Sarkis, Henning, Kürster, Trifonov,
  Zechmeister, Tal-Or, Anglada-Escudé, Hatzes, Lafarga, Dreizler, Ribas,
  Caballero, Reiners, Mallonn, Morales, Kaminski, Aceituno, Amado, Béjar,
  Hagen, Jeffers, Quirrenbach, Launhardt, Marvin, \& Montes}]{Sarkis_2018}
Sarkis, P., Henning, T., Kürster, M., {et~al.} 2018, The Astronomical Journal,
  155, 257, \dodoi{10.3847/1538-3881/aac108}

\bibitem[{Seager {et~al.}(2007)Seager, Kuchner, Hier-Majumder, \&
  Militzer}]{Seager_2007}
Seager, S., Kuchner, M., Hier-Majumder, C.~A., \& Militzer, B. 2007, The
  Astrophysical Journal, 669, 1279, \dodoi{10.1086/521346}

\bibitem[{{Seager} \& {Mall{\'e}n-Ornelas}(2003)}]{2003ApJ...585.1038S}
{Seager}, S., \& {Mall{\'e}n-Ornelas}, G. 2003, \apj, 585, 1038,
  \dodoi{10.1086/346105}

\bibitem[{{Selsis} {et~al.}(2007){Selsis}, {Kasting}, {Levrard}, {Paillet},
  {Ribas}, \& {Delfosse}}]{Selsis2007A}
{Selsis}, F., {Kasting}, J.~F., {Levrard}, B., {et~al.} 2007, \aap, 476, 1373,
  \dodoi{10.1051/0004-6361:20078091}

\bibitem[{{Selsis, F.} {et~al.}(2011){Selsis, F.}, {Wordsworth, R. D.}, \&
  {Forget, F.}}]{Selsis_2011}
{Selsis, F.}, {Wordsworth, R. D.}, \& {Forget, F.} 2011, A\&A, 532, A1,
  \dodoi{10.1051/0004-6361/201116654}

\bibitem[{Showman \& Malhotra(1997)}]{SHOWMAN_1997}
Showman, A.~P., \& Malhotra, R. 1997, Icarus, 127, 93,
  \dodoi{https://doi.org/10.1006/icar.1996.5669}

\bibitem[{{Sleep} \& {Zahnle}(2001)}]{Sleep2001}
{Sleep}, N.~H., \& {Zahnle}, K. 2001, \jgr, 106, 1373,
  \dodoi{10.1029/2000JE001247}

\bibitem[{{Smith} {et~al.}(2018){Smith}, {Fratanduono}, {Braun}, {Duffy},
  {Wicks}, {Celliers}, {Ali}, {Fernandez-Pa{\~n}ella}, {Kraus}, {Swift},
  {Collins}, \& {Eggert}}]{Smith2018}
{Smith}, R.~F., {Fratanduono}, D.~E., {Braun}, D.~G., {et~al.} 2018, Nature
  Astronomy, 2, 452, \dodoi{10.1038/s41550-018-0437-9}

\bibitem[{Sotin {et~al.}(2007)Sotin, Grasset, \& Mocquet}]{SOTIN_2007}
Sotin, C., Grasset, O., \& Mocquet, A. 2007, Icarus, 191, 337,
  \dodoi{https://doi.org/10.1016/j.icarus.2007.04.006}

\bibitem[{{Tobie, G.} {et~al.}(2019){Tobie, G.}, {Grasset, O.}, {Dumoulin, C.},
  \& {Mocquet, A.}}]{Tobie_2019}
{Tobie, G.}, {Grasset, O.}, {Dumoulin, C.}, \& {Mocquet, A.} 2019, A\&A, 630,
  A70, \dodoi{10.1051/0004-6361/201935297}

\bibitem[{{Turbet, Martin} {et~al.}(2018){Turbet, Martin}, {Bolmont, Emeline},
  {Leconte, Jeremy}, {Forget, Fran\c{c}ois}, {Selsis, Franck}, {Tobie,
  Gabriel}, {Caldas, Anthony}, {Naar, Joseph}, \& {Gillon,
  Micha\"el}}]{Turbet2018}
{Turbet, Martin}, {Bolmont, Emeline}, {Leconte, Jeremy}, {et~al.} 2018, A\&A,
  612, A86, \dodoi{10.1051/0004-6361/201731620}

\bibitem[{{Turner, Jake D.} {et~al.}(2021){Turner, Jake D.}, {Zarka, Philippe},
  {Grie\ss{}meier, Jean-Mathias}, {Lazio, Joseph}, {Cecconi, Baptiste}, {Emilio
  Enriquez, J.}, {Girard, Julien N.}, {Jayawardhana, Ray}, {Lamy, Laurent},
  {Nichols, Jonathan D.}, \& {de Pater, Imke}}]{Turner_2021}
{Turner, Jake D.}, {Zarka, Philippe}, {Grie\ss{}meier, Jean-Mathias}, {et~al.}
  2021, A\&A, 645, A59, \dodoi{10.1051/0004-6361/201937201}

\bibitem[{{Underwood} {et~al.}(2003){Underwood}, {Jones}, \&
  {Sleep}}]{Underwood2003}
{Underwood}, D.~R., {Jones}, B.~W., \& {Sleep}, P.~N. 2003, International
  Journal of Astrobiology, 2, 289, \dodoi{10.1017/S1473550404001715}

\bibitem[{{Unterborn} {et~al.}(2018){Unterborn}, {Desch}, {Hinkel}, \&
  {Lorenzo}}]{UnterbornEXOPLEX2018}
{Unterborn}, C.~T., {Desch}, S.~J., {Hinkel}, N.~R., \& {Lorenzo}, A. 2018,
  Nature Astronomy, 2, 297, \dodoi{10.1038/s41550-018-0411-6}

\bibitem[{Unterborn {et~al.}(2016)Unterborn, Dismukes, \&
  Panero}]{Unterborn_2016}
Unterborn, C.~T., Dismukes, E.~E., \& Panero, W.~R. 2016, The Astrophysical
  Journal, 819, 32, \dodoi{10.3847/0004-637X/819/1/32}

\bibitem[{Unterborn {et~al.}(2022)Unterborn, Foley, Desch, Young, Vance,
  Chiffelle, \& Kane}]{Unterborn_2022}
Unterborn, C.~T., Foley, B.~J., Desch, S.~J., {et~al.} 2022, The Astrophysical
  Journal Letters, 930, L6, \dodoi{10.3847/2041-8213/ac6596}

\bibitem[{Valencia {et~al.}(2007)Valencia, Sasselov, \&
  O’Connell}]{Valencia_2007}
Valencia, D., Sasselov, D.~D., \& O’Connell, R.~J. 2007, The Astrophysical
  Journal, 665, 1413, \dodoi{10.1086/519554}

\bibitem[{{van den Berg} {et~al.}(2019){van den Berg}, {Yuen}, {Umemoto},
  {Jacobs}, \& {Wentzcovitch}}]{vandenBerg2019}
{van den Berg}, A.~P., {Yuen}, D.~A., {Umemoto}, K., {Jacobs}, M.~H.~G., \&
  {Wentzcovitch}, R.~M. 2019, \icarus, 317, 412,
  \dodoi{10.1016/j.icarus.2018.08.016}

\bibitem[{{Varela} {et~al.}(2018){Varela}, {R{\'e}ville}, {Brun}, {Zarka}, \&
  {Pantellini}}]{Varela2018}
{Varela}, J., {R{\'e}ville}, V., {Brun}, A.~S., {Zarka}, P., \& {Pantellini},
  F. 2018, \aap, 616, A182, \dodoi{10.1051/0004-6361/201732091}

\bibitem[{{Vazan} {et~al.}(2013){Vazan}, {Kovetz}, {Podolak}, \&
  {Helled}}]{Vazan2013}
{Vazan}, A., {Kovetz}, A., {Podolak}, M., \& {Helled}, R. 2013, \mnras, 434,
  3283, \dodoi{10.1093/mnras/stt1248}

\bibitem[{{Vogt} {et~al.}(1994){Vogt}, {Allen}, {Bigelow}, {Bresee}, {Brown},
  {Cantrall}, {Conrad}, {Couture}, {Delaney}, {Epps}, {Hilyard}, {Hilyard},
  {Horn}, {Jern}, {Kanto}, {Keane}, {Kibrick}, {Lewis}, {Osborne},
  {Pardeilhan}, {Pfister}, {Ricketts}, {Robinson}, {Stover}, {Tucker}, {Ward},
  \& {Wei}}]{1994SPIE.2198..362V}
{Vogt}, S.~S., {Allen}, S.~L., {Bigelow}, B.~C., {et~al.} 1994, in Society of
  Photo-Optical Instrumentation Engineers (SPIE) Conference Series, Vol. 2198,
  Instrumentation in Astronomy VIII, ed. D.~L. {Crawford} \& E.~R. {Craine},
  362, \dodoi{10.1117/12.176725}

\bibitem[{Wandel(2018)}]{Wandel_2018}
Wandel, A. 2018, The Astrophysical Journal, 856, 165,
  \dodoi{10.3847/1538-4357/aaae6e}

\bibitem[{Wandel(2023)}]{Wandel_2023Extended}
---. 2023, The Astronomical Journal, 166, 222, \dodoi{10.3847/1538-3881/ad0045}

\bibitem[{{Wang} {et~al.}(2022){Wang}, {Shen}, {Luo}, {Yang}, {Gai}, {Tang},
  {Wang}, {Qin}, {Han}, \& {Rong}}]{2022ApJS..258....9W}
{Wang}, L.-L., {Shen}, S.-Y., {Luo}, A.~L., {et~al.} 2022, \apjs, 258, 9,
  \dodoi{10.3847/1538-4365/ac3241}

\bibitem[{{Wolf} {et~al.}(2019){Wolf}, {Kopparapu}, {Airapetian}, {Fauchez},
  {Guzewich}, {Kane}, {Pidhorodetska}, {Way}, {Abbot}, {Checlair}, {Davis},
  {Genio}, {Dong}, {Eggl}, {Fleming}, {Fujii}, {Haghighipour}, {Heavens},
  {Henning}, {Kiang}, {L{\'o}pez-Morales}, {Lustig-Yaeger}, {Meadows},
  {Reinhard}, {Rugheimer}, {Schwieterman}, {Shields}, {Sohl}, {Turbet}, \&
  {Wordsworth}}]{Wolf_2019}
{Wolf}, E., {Kopparapu}, R., {Airapetian}, V., {et~al.} 2019, Astro2020:
  Decadal Survey on Astronomy and Astrophysics, 2020, 177,
  \dodoi{10.48550/arXiv.1903.05012}

\bibitem[{{Wunderlich, Fabian} {et~al.}(2019){Wunderlich, Fabian}, {Godolt,
  Mareike}, {Grenfell, John Lee}, {St\"adt, Steffen}, {Smith, Alexis M. S.},
  {Gebauer, Stefanie}, {Schreier, Franz}, {Hedelt, Pascal}, \& {Rauer,
  Heike}}]{Wunderlich_2019}
{Wunderlich, Fabian}, {Godolt, Mareike}, {Grenfell, John Lee}, {et~al.} 2019,
  A\&A, 624, A49, \dodoi{10.1051/0004-6361/201834504}

\bibitem[{Yang {et~al.}(2014)Yang, Boué, Fabrycky, \& Abbot}]{Yang_2014}
Yang, J., Boué, G., Fabrycky, D.~C., \& Abbot, D.~S. 2014, The Astrophysical
  Journal Letters, 787, L2, \dodoi{10.1088/2041-8205/787/1/L2}

\bibitem[{Yang {et~al.}(2013)Yang, Cowan, \& Abbot}]{Yang_2013}
Yang, J., Cowan, N.~B., \& Abbot, D.~S. 2013, The Astrophysical Journal
  Letters, 771, L45, \dodoi{10.1088/2041-8205/771/2/L45}

\bibitem[{{Zacharias} {et~al.}(2013){Zacharias}, {Finch}, {Girard}, {Henden},
  {Bartlett}, {Monet}, \& {Zacharias}}]{Zacharis_2012_catalogue}
{Zacharias}, N., {Finch}, C.~T., {Girard}, T.~M., {et~al.} 2013, \aj, 145, 44,
  \dodoi{10.1088/0004-6256/145/2/44}

\bibitem[{{Zeng} {et~al.}(2016){Zeng}, {Sasselov}, \& {Jacobsen}}]{Zeng2016}
{Zeng}, L., {Sasselov}, D.~D., \& {Jacobsen}, S.~B. 2016, \apj, 819, 127,
  \dodoi{10.3847/0004-637X/819/2/127}

\bibitem[{Zeng \& Seager(2008)}]{Zeng_2008}
Zeng, L., \& Seager, S. 2008, Publications of the Astronomical Society of the
  Pacific, 120, 983, \dodoi{10.1086/591807}

\end{thebibliography}
\bibliographystyle{aasjournal}

\end{document}